 \documentclass[final,5p,times,authoryear]{elsarticle}

\usepackage{amssymb,amsmath}
\usepackage{xcolor}

\usepackage{soul}
\usepackage{pstool}

\usepackage{booktabs}
\usepackage{multirow} 
\usepackage{array}
\usepackage{caption}
\usepackage{soul}

\journal{Annual Reviews in Control}

\begin{document}

\begin{frontmatter}



\title{Non-integer (or fractional) power model \\to represent the complexity of a viral spreading: application to the COVID-19}


\author[IMS]{Alain Oustaloup}\ead{alain.oustaloup@ims-bordeaux.fr}
\author[IMB]{Fran{\c{c}}ois Levron}\ead{f.levron@wanadoo.fr}
\author[IMS]{St\'ephane Victor \corref{cor1}}\ead{stephane.victor@ims-bordeaux.fr}
\author[GIPSA]{Luc Dugard} \ead{luc.dugard@gipsa-lab.grenoble-inp.fr}

\address[IMS]{Univ. Bordeaux, CNRS, IMS UMR 5218, Bordeaux INP/ENSEIRB-MATMECA, \\
351 Cours de la Lib\'eration, 
33405 Talence CEDEX, France}
\address[IMB]{Univ. Bordeaux, CNRS, IMB UMR 5251, Bordeaux INP/ENSEIRB-MATMECA, \\
351 Cours de la Lib\'eration, 
33405 Talence CEDEX, France}
\address[GIPSA]{Univ. Grenoble Alpes, CNRS, Grenoble INP, GIPSA-Lab, 38000 Grenoble, France}

\cortext[cor1]{Corresponding author}

\begin{abstract}
This article proposes a very simple deterministic mathematical model, which, by using a power-law, is a \emph{non-integer power model} (or \emph{fractional power model (FPM)}). Such a model, in non-integer power of time, namely $t^m$ up  to constants, enables representing the totality of the contaminated individuals at each day, with a good precision, thus expressing the interest of this tool for time series. Despite being enriched with knowledge through an internal structure based on a geometric sequence ``with variable ratio'', the model (in its non-integer representation) has only three parameters, among which the non-integer power, $m$, that determines on its own, according to its value, an aggravation or an improvement of the viral spreading. Its simplicity comes from the power-law, $t^m$, which simply expresses the singular dynamics of the operator of non-integer differentiation or integration, of high parametric compactness, that governs diffusion phenomena and, as shown in this article, the spreading phenomena by contamination. The representativity of the proposed model is indeed validated with the official data of French Ministry of Health on the COVID-19 spreading, notably the time series of the contaminations and the hospitalizations. Used in prediction, the model well enables justifying the choice of a lockdown, without which the spreading would have highly worsened. Its predictivity is validated by verified predictions in lockdown and vaccination phases, and even for the vaccination itself; its simplicity enables a very simple implementation of the prediction technique.
The comparison of this model in $t^m$ with two known models having the same number of parameters, well shows that its representativity of the real data is better or more general. Finally, in a more fundamental context and particularly in terms of complexity and simplicity, a self-filtering action enables showing the compatibility between the \emph{internal complexity} that the internal structure and its stochastic behavior present, and the \emph{global simplicity} that the model in $t^m$ offers in a deterministic manner: it is true that the non-integer power of a power-law,that an \emph{internal dynamics dispersion} can justify, is well a marker of complexity, and this, beyond viral spreading phenomena.
\end{abstract}



\begin{keyword}
COVID-19, viral spreading, modeling, prediction; {{time series}}; fractional (or non-integer) power model (FPM), a FPM model as a convexity or concavity model; power-law, internal dynamics dispersion, non-integer (or fractional) differentiation or integration, self-filtering internal structure. 
\end{keyword}

\end{frontmatter}



\section{Introduction}\label{sec:Intro}
\subsection{On our knowledge of the existing}

The most common approach for modeling an epidemic is a behavior model, originally proposed by \cite{Kermack1927}. The population is divided into several categories and mathematical rules dictate how many people move from one category to another. 

First, everyone is \emph{Susceptible} (S) to become contaminated.
Then some people become \emph{Infected} (I) individuals who have been infected and are capable of infecting susceptible individuals. Finally, individuals, who have been infected enter the \emph{Removed} (R) compartment, through either recovery or death. These SIR  models can be improved by including \emph{Exposed} (E) people but not yet contagious, leading to SEIR models. If post-recovery immunity is temporary, recovered people can go back to S, thus leading to SEIRS models. The equations that determine how people move from one category to the next depend on a wide variety of parameters drawn from biology, behavior, politics, economy, weather, and more. Some of the extended SEIRS models can be found in \cite{Satsuma2004,Kwuimy2020,He2020,Ivorra2020,DellAnna2020,Demongeot2020,Guan2020,Efimov2021}. 
A model linked to SIR models is the time-modulated Hawkes process (\cite{Garetto2021}).
Other extensions of the SEIRS models have been carried out by using fractional (or non-integer) order derivatives  (\cite{Xu2020,Lu2020,Arfan2020}), as the non-integer operator is known to well model propagation phenomena such as biological systems, thermal diffusion systems, electro-che\-mical ones (fuel cells or batteries), etc.

Parameter regression algorithms such as used in system identification, can also be used for monitoring and forecasting the behavior of epidemics such as dynamic harmonic regression algorithms (\cite{Young2021}), the combination of a macroscopic model and a local model (\cite{Scharbarg2020}).

Other kinds of spreading models exist in the literature to avoid sorting out people per categories, namely data-driven mo\-dels. Black box models are used but the complexity and huge number of parameters render these models disconnected from reality and beyond understanding. On one hand, these models, as being more tied to data, can provide more accurate predictions in the short-term, but on the other hand no guarantee is proven for long-term predictions. Some data-driven models use neural networks, deep learning, machine learning, YYG model\footnote{https://covid19-projections.com developed by Youyang Gu}, particle swarm optimization, etc. (see for example \cite{Liu2020,Huang2020,ZhaoShi2020,De-Leon2020}). 

Infection, death, and hospitalization numbers affect their usefulness not only as inputs for the model but also as outputs. The true number of infections is hard to evaluate when not eve\-ryone is tested. Deaths are countable, but they are the consequence of infections after some time delay. Therefore, other kinds of models have emerged by proposing transmission, networking and control measures to reduce the epidemic spreading such as the agent-based models, the sliding mode control (\cite{Rohith2020}), mask wearing effects (\cite{Li2020}), human mobility control (\cite{Iacus2020}), model predictive control (\cite{Carli2020}), theoretical roots of COVID dynamics (\cite{LeMehaute2021}), lockdown effect, etc. (\cite{HuangJianzhe2020,De-Visscher2020, Chen2020,Atangana2020,Tuan2020}).

\cite{Ferguson2005} have developed a stochastic model for the spread of infectious diseases, which also takes into account the geographic spreading. Simple models such as SEIRS ones are not sufficiently accurate to evaluate the spreading of an epidemic. There is a need for simple models that are sufficiently accurate for spreading predictions, so that quick and real-time decisions can be made on how to respond to epidemics.
The Belgian \cite{Verhulst1838} established a model of dynamical behavior of a population, named logistic growth model. 
Many studies use this model with three parameters for modeling the spreading of COVID-19 (\cite{Zhou2020,Kyurkchiev2020,Zhao2020}).
There are several possible generalizations of the Verhulst model such as proposed by \cite{Roosa2020} who give two well-known generalizations with four parameters: the generalized logistical growth model (\cite{Viboud2016,Ganyani2018}) and the generalized Richards model (\cite{Richards1959,Wang2012,Wu2020}).

Some contributions use a power-law. Whereas this power-law is a function of a distance (\cite{Meyer2014}) or of a contact duration (\cite{Duan2015}), that concerns a contamination probability, in our article, power-law is a time function that concerns the deterministic evolution of the total number of the contaminated. Besides, the change from the stochastic to the deterministic, via the change to the global that implies a smoothing (or integration) operation, does not seem incompa\-tible with maintaining a time form in power-law.

\subsection{On the specific contributions of the article}\label{subsec:contributions}
The specific contributions of the article turn on a very simple deterministic mathematical model, in non-integer power of time, namely $t^m$ up to constants, which enables representing, at each day, the totality of the contaminated individuals (which, from now on, will be named contaminated); and this, with an absolutely satisfactory precision in comparison with the complexity of contamination phenomena. 
{{More generally, through its non-integer power, $m$, this model (said to be non-integer) constitutes a very interesting tool to simply and precisely represent times series.\label{Rev2:TimeSeries}
The model, $A+Bt^m$, is characterized by three parameters, the non-integer power, $m$, as well as two constants, $A$ and $B$, one being simply additive and the other  being multiplicative of $t^m$. }}
The non-integer power, $m>0$, determines on its own a viral sprea\-ding that worsens or gets better according to whether $m$ is greater or lesser than one: indeed, the spreading graphical representation presents, for $m$ greater than one, a convexity that expresses a spreading increasing more and more, therefore a viral aggravation ; whereas for $m$ lesser than one, the curve presents a concavity that expresses a spreading increasing less and less, therefore a viral improvement. {{Note that $m$ plays an analogous role as the one of $R_0$ in the SIR model, $R_0$ being the basic reproduction number (\ref{annex:B}). }}

The simplicity of the model in $t^m$ (FPM) is due to the fact that $t^m$ simply expresses a characteristic response, namely the step response, therefore the dynamics, of the non-integer differentiation or integration operator. Applicable in real-time by means of Oustaloup's approximation (\cite{Oustaloup2000}), this operator of non-integer order, $m$ in integration, is known for its parametric compactness and its capacity to govern complex phenomena (\cite{Oustaloup1995a,Oustaloup2014}), among which the diffusion phenomena and, as shown in this article, the contamination spreading phenomena. {{Thus, through the non-integer power, $m$, of the model in $t^m$, this article is a contribution to the unification of  diffusion phenomena in physics and viral spreading phenomena in epidemiology, and this, more precisely at medium times, therefore at medium frequencies. It is true that in nature,  and more generally in reality, a non-integer behavior can only occur in a medium frequency range. That amounts to saying that the model in $t^m$ does not have the vocation to be representative of reality, at short times, which correspond to high frequencies and at long times, which correspond to low frequencies.  }}

In order not to limit this model to a simple behavior model, it is enriched by knowledge (of countable nature) through an \emph{internal structure}, in accordance with the daily recording of the new contaminated, and founded on a geometric sequence ``with variable ratio'', since the internal structure expression is none other than the sum of the $k+1$ first terms of this sequence (sections \ref{sec:sequence_elem} and \ref{sec:WhysuiteGeomVar}). For each of the values of $k$, this sum enables constructing point by point (in this case day after day) a non-integer time function representing the evolution of the total number of the new contaminated: this three parameter function (section \ref{sec:TimeModelSpreading}) has the same form (therefore the same degree, $m$) as the model in $t^m$, even in its complete version that represents the evolution of the number of all the contaminated, new and former ones (section \ref{subsec:ModelContamine}).



The internal structure, through which the model wins in knowledge, very simply enables formalizing the process of daily census of the new contaminated, and to represent day after day the whole of the new contaminated. To that effect, the internal structure is conceived for introducing ratios, $q_k$, between new contaminated  between two consecutive days $k-1$ and $k$ (sections \ref{sec:sequence_elem} and \ref{sec:WhysuiteGeomVar}), then used without seeking to capture or represent the unpredictable, therefore the hazards of reality, in this case the random fluctuations of these ratios, which occur from day to day in reality given its complexity. Adopted for obvious reasons of simplicity, such a strategy is all the more justified as the fluctuations in question, besides interpretable as an internal noise, have only a negligible effect on the dynamics to be modelled, slower, as being expressed on se\-veral days. What follows is liable to precise the content of these words, and to reinforce the idea of not systematically seeking to represent the whole complexity of reality, particularly all these random fluctuations.

The internal structure of the model in $t^m$ presents a remarkable property of filtering, through a \emph{double effect} inherent to the form itself of its expression such as conditioned by the geometric sequence ``with variable ratio'', namely the sum of the products of the different ratios, $q_1,\ldots,\, q_k$ (sections \ref{sec:sequence_elem} and \ref{sec:WhysuiteGeomVar}): 
\begin{itemize}
\item the first effect of filtering finds its cause in the terms of the sequence, that is to say the products of the different ratios, products that can filter by themselves the fluctuations of these ratios to only present residual fluctuations;
\item the second effect of filtering finds its cause in the sum of the sequence terms, namely the non weighted sum of the products of the different ratios, a sum that therefore presents a (discrete) integral action that can filter at its turn the residual fluctuations of the products.
\end{itemize}

The specificity of such a property, that confers to the internal structure a \emph{self-filtering} character with double effect, stands out by the joint effect of a sum and of products, thus overtaking the filtering effect of a simple integrator, naturally limited to the one of a sum. 

This \emph{self-filtering double effect} is validated in section \ref{subsec:ValidationAction_tm}, through real curves (figure \ref{fig:qk_uk_sk}) which prove that the stochastic behavior due to the random fluctuations of the $q_k$, has globally no incidence or practically not, notably on the sum, $s_k$, of the new contaminated at day $k$. The modeling of this sum (or more precisely its result), which is an objective in epidemiology and then in this article,  therefore needs only a \emph{deterministic model}, in this case the model in $t^m$, thus expressing the compatibility between \emph{internal complexity} and \emph{global simplicity}.

In other words, this self-filtering double effect liable to globally denoise the internal structure of the model in $t^m$, well expresses a coherent physical phenomenon, in the sense that the model in $t^m$ is not noisy, even if it is noisy internally. This phenomenon assuredly is likely to remove the paradox between the simplicity of a model in $t^m$ and the complexity of a noisy reality, thus reinforcing the interest of the parametric compactness of a model in $t^m$ in the modeling of complex systems and phenomena (including among others those of finance (see \ref{App:AnnexePower-law})).

\subsection{On the complexity that the internal structure covers: an explanation for the global behavior in $t^m$}\label{subsec:ComplexityInternalStructure}
Concerning the link between the complex phenomena and the non-integer, in \cite{Oustaloup2014}, \emph{diversity} that participates on complexity, is associated with the non-integer through multiplicity and difference that it presents whatever its form. 
Thus, under this angle, what about diversity in our study case, especially since the contamination phenomena being natural, they cannot escape from diversity. As the new contaminated recorded at day $k$ result from the various contaminations that occur between days $k-1$ and $k$, the internal structure that lists all new contaminated through a daily recording, thus finds  its origin in a complexity of another nature and of another degree. The internal structure indeed results from a diversity form that well responds to multiplicity and difference, namely \emph{a multitude of elements likely to change state in different time and in diffe\-rent number}: the difference associated with multiplicity thus turns, both, on time and on the number of state changes. 
More precisely, if each state change is defined by the change from a state 0 to a state 1, and if state 0 is a  non-contamination state and state 1 a contamination state (the elements then being individuals), the diversity form in question corresponds to the spreading process of a viral contamination. The state changes between two consecutive days $k-1$ and $k$ correspond to new contaminated at day $k$, and the stage changes between days $-1$ and $k$ correspond to the sum (or total number) of the new contaminated at day $k$. Given this phenomenological analysis and the representation in $t^m$ of the real evolution of the total number of the new contaminated, as validated by identification in section \ref{sec:Validation_tm_COVID19}, the stage changes, different in time and in number, are therefore made according to a global behavior which is characteristic of the non-integer.

Actually, such a phenomenon can find an explanation in a dispersion of \emph{internal dynamics} of contamination, since it is now well known that a dispersion of dynamics is at the origin of a non-integer behavior (\cite{Oustaloup2000,Oustaloup2014}). To highlight such a dispersion, an analysis can be led from the state changes that occur between days $0$ and $k$, and that condition the total number evolution of the new contaminated at day $k$, that the constant, $u_0$, initializes with the new contaminated recorded at day $0$ (namely the contaminated between days $-1$ and $0$). It suffices to analyze, in dynamics terms, the effects of a joint dispersion of the times and the numbers of the state changes: if the state changes are distant in time and low in number, they generate slow dynamics; if the state changes are close in time and high in number, they generate fast dyna\-mics. Between these two cases, all possible combinations in time and number are liable to generate dynamics of interme\-diary rapidities or, more precisely, more or less slow dynamics, medium rapidity dynamics, and more or less fast dynamics. Given that the global behavior results from the sum of all state changes, it is possible to sum all dynamics in an ordered manner, by going, for example, from the slowest to the fastest dynamics. Thus corresponding to a \emph{regular growth of rapidities}, such a sum is not without evoking the \emph{recursive distribution of aperiodic dyna\-mics} which is at the origin of a non-integer behavior (\cite{Oustaloup1995a,Oustaloup2014}). For the step response of an integrator of non-integer order, $m$, which is at stake in this article, the distribution is the one of concave dynamics for $m<1$ and of convex dynamics for $m>1$: such a distribution can be obtained from a partial fraction decomposition of the synthesis transmittance of a integrator of order $m$ between 0 and 1 (\cite{Trigeassou2019b}); the transmittance is then founded on a \emph{recursive distribution of strictly alternated real zeros and poles} (\cite{Oustaloup1995a,Oustaloup2000}).    

{

\subsection{A global behavior in $t^m$ beyond spreading phenomena by contamination}\label{subsec:GlobalBehaviorProPhenContamination}
The non-integer behaviors are relatively numerous, insofar as they can already occur in complex systems and phenomena, whose phenomenological analysis can justify \emph{regularly distri\-buted internal dynamics}: these dynamics can go from the slo\-west one to the fastest one with a regular growth of the rapidities. Such dynamics are even more frequent as their origin is different according to the process nature that generates them. They are indeed not generated in the same way: the dispersion of the internal dynamics can result from different time constants, stemming from different resistances and capacitances and/or  different action level, which is notably the case for the study of the famous porous dyke and for the hydropneumatic version of the CRONE suspension that results from it (\cite{Oustaloup2014}); but this type of dynamics dispersion  can also result from state changes in different time  and in different number, consecutive to diverse causes. Among these causes, one can cite, not only the contaminations (section \ref{subsec:ComplexityInternalStructure}), but also the vaccinations (section \ref{subsec:PredictionVaccinationEllememe}), the births, the deceased, ... and even some financial movements (\ref{App:AnnexePower-law}). 

That amounts to saying that a global behavior in $t^m$ is not limited to viral spreading phenomena. That is why a model in $t^m$ constitutes a good modeling tool to represent complexity, and this, beyond that of a viral spreading.  
}

\section{Idea precision of the article in a systemic approach}
\label{sec:SystemicApproach}
Any real system naturally presents an internal organization that achieves it  or synthesizes it, for example physically or chemically. This internal organization or structure, conditions the system characteristic quantities. In a systemic approach coming under system theory, these quantities are represented by internal and external variables that depend on structural parameters and that generally figure in a functional diagram. 

So, coming within such an approach, the idea of the article is to propose a global model in power-law, of the form $A+Bt^m$, likely to represent the dynamics of complex phenomena, and to equip it with an internal and explicative structure, in conformity with a daily reality (of countable nature).

More precisely, concerning the nature of such a model, the aim is to introduce an \emph{internal structure} (or \emph{explicative model})
\begin{itemize}
\item which, through each of its terms (see hereafter), well represents the daily recording of the new contaminated 
\item and which, through the whole of its terms (therefore globally), gives the real evolution of the total number of the new contaminated, namely the evolution that the model is supposed to represent.
\end{itemize}

The internal structure thus enables formalizing the daily recording process of the new contaminated and to represent day after day the number of all the new contaminated. It is mathe\-matically defined by the sum, $s_k$, of the $k+1$ first terms of a geometrical sequence ``with variable ratio'' such as defined in section \ref{sec:sequence_elem}, namely:
\[
u_0+u_0q_1+ u_0q_1q_2 + \ldots+ u_0 q_1\ldots q_k, \, \text{with}\,  u_0,\, q_1,\ldots, \, q_k >0,
\]
$q_k$ being the ratio of the new contaminated between two consecutive days $k-1$ and $k$.

While the sequence general term, $u_k=u_0q_1\ldots q_k$, represents the number of the new contaminated at day $k$ ($u_0$ being relative to day $0$), the sum, $s_k$, associated with the sequence, represents the sum of the new contaminated recorded from day $0$ to day $k$ or, more simply, the sum of the new contaminated at day $k$ (section  \ref{sec:WhysuiteGeomVar}).

Through the products of the $q_k$ and the sum of the $u_k$ of the internal structure expression, $s_k$, that  confer a self-filtering double effect to the structure (section \ref{subsec:ValidationAction_tm}), the random fluctuations that the different ratios, $q_k$, can present, do not practically disturb the sum, $s_k$, or more precisely, the result of the sum. So, the model, that is supposed to represent the evolution of this sum (or more precisely its result), can be a model essentially deterministic, thus justifying the nature of the proposed model in the article. If the internal structure therefore enables explaining the deterministic nature of the global behavior, it also enables to explain the form in $t^m$ of this behavior: with its countable way to capture, day after day, a noisy reality (including recording errors), this type of structure covers a complexity that explains its global behavior in $t^m$, through a diversity form that leads to a \emph{regular distribution of internal dynamics} (section \ref{subsec:ComplexityInternalStructure}).

Stemming from the behavior of the fractional integration operator, which is particularly important in modeling complex systems and phenomena, the proposed model is well a deterministic model. It is, mathematically, defined by an affine function of the power-law, $t^m$, namely:
\[
u_0 (1+ct^m) \quad \text{with} \quad c,m>0 ;
\]
this expression with three parameters, $u_0$, $c$ and $m$, where $m$ is a parameter ``of high level'', is presented as an argued hypothesis in section \ref{sec:TimeModelSpreading}, then validated with the official data on COVID-19 in section \ref{sec:Validation_tm_COVID19}.

In the case where the model expression well represents the evolution of the total number of the new contaminated, that is to say the evolution of the sum, $s_k$, there is then a relation of equality between the internal structure and the model expressions. This equality relation thus determines an equation such as established in section \ref{sec:TimeModelSpreading}: this equation can be reduced to equation \eqref{eq:first}, then equation \eqref{eq:first1}, from which the different ratios, $q_k$, can be determined (relation \eqref{eq:first5}); the $q_k$ in question are the theoretical $q_k$ that smooth the real $q_k$ and that thus characterizes the average behavior of the internal structure, to which the stochastic behavior is superposed (section \ref{subsec:Validation_tm}).

Finally, in the internal and global approach that the article develops, the self-filtering action of the internal structure, which assumes random fluctuations of zero mean, proves to be well justified, since the results of section \ref{subsec:Validation_tm}  show (up to errors) that the random fluctuations of the real $q_k$, around the theoretical $q_k$ inherent to the model, present a zero mean. Besides, this mean nullity reinforces the model validity, in the sense that the model, which well represents the global behavior, also permits to obtain, in internal, theoretical $q_k$ that smooth the real $q_k$  so that the fluctuation mean is nil.

\section{Geometric sequence with variable ratio and associated sum}
\label{sec:sequence_elem}


Such as lead in this article, establishing a viral spreading model via an internal structure, calls, so it seems, 
a targeted mathematical recall as regards sequences, in order to better understand the geometric sequence ``with variable ratio'', introduced here to define the internal structure.

In mathematics, a \emph{sequence}, denoted $\left( u_k\right)_{k\geq 0}$, is written as the \emph{sequence of terms}:
\[
u_0,\, u_1, \, u_2,\, \ldots,\, u_k, \, \ldots;
\]
of positive or null subscript $k$ which indicates the rank, the \emph{gene\-ral term (of rank $k$)}, $u_k$, defines 
the sequence on its own.
The sequence is \emph{increasing} if, for all $k$, the term of rank $k+1$ is greater than the term of rank $k$, namely $u_{k+1}>u_k$. It is \emph{decreasing} if, for all $k$, the term of rank $k+1$ is lesser than the term of rank $k$, namely $u_{k+1}<u_k$.

Concerning the sum of the terms of the so defined sequence, in particular the first ones, the sum, $s_k$, of the $k+1$ first terms of the sequence $\left( u_k\right)_{k\geq 0}$, is expressed according to the sum of terms:
\[
s_k = u_0+u_1+\ldots+u_k.
\]

For the most initiated ones, this sum $s_k$, associated with the sequence $\left( u_k\right)_{k\geq 0}$, is not without recalling the general term, $s_k$, of the \emph{sequence} of the partial sums of the terms of $\left( u_k\right)_{k\geq 0}$, that defines the \emph{series}, $\left( s_k\right)_{k\geq 0}$: indeed, \emph{if a sequence is a sequence of terms, a series is a sequence of term sums}.

An \emph{arithmetic sequence} (or \emph{arithmetic progression}) is charac\-terized by a \emph{{constant} difference}, $r$, between two consecutive terms, namely:
\[
u_{k+1}-u_k = r \quad \forall k\geq 0 ;
\]
$r$ is called \emph{common difference} of the sequence.

The term $u_{k+1}$ with $k\geq 0$, can be written under the form of a \emph{recurrence relation}, namely 
\[
u_{k+1} = u_k+r,
\]
which expresses that each term of the sequence is the \emph{sum} of the precedent and the common difference. 
%

%
A \emph{geometric sequence} (or \emph{geometric progression}) is charac\-terized by a \emph{{constant} ratio}, $q$, between two consecutive terms, namely:
\[
\frac{u_{k+1}}{u_k} = q \quad \forall k\geq 0  \quad \text{and} \quad u_0 \neq 0;
\]
$q$ is called \emph{common ratio} of the sequence.

The term $u_{k+1}$ with $k\geq 0$, can be written under the form of a \emph{recurrence relation}, namely
\[
u_{k+1} = u_k\,q,
\]
which expresses that each term of the sequence is the \emph{product} of the precedent with the common ratio. 
%
%



Whilst in a simple geometric sequence, the ratio between two consecutive terms is constant, in the geometric sequence considered here, said \emph{``with variable ratio''}, the ratio is no longer constant, since it is different according to the rank $k$ (as being dependent on it), \emph{thus varying with the rank}, namely:
\[
\frac{u_{k+1}}{u_k} = q_{k+1} \quad \forall k\geq 0  \quad \text{and} \quad u_0 \neq 0,
\]
the term $u_{k+1}$, $k\geq 0$, being then defined by the recurrence relation:
\[
u_{k+1} = u_k q_{k+1}.
\]
Thus:
\begin{align*}
u_1& = u_0 \, q_1,\\
u_2 &= u_1\, q_2= u_0\, q_1\, q_2,\\
u_3 &= u_2\, q_3=u_0\, q_1\, q_2\, q_3,\dots,
\end{align*} 
the general term of the sequence then being:
\[
u_k = u_0 q_1\ldots q_k \quad \forall k\geq 1.
\]

The sum $s_k$ associated with the so defined geometric sequence ``with variable ratio'', is given by the sum of terms:
\[
s_k = u_0+u_1+u_2+\ldots+ u_k \quad \text{with}\quad u_k = u_0 q_1\ldots q_k, k\geq 1,
\]
namely:
\[
s_k = u_0(1+q_1+q_1q_2+\ldots+ q_1\ldots q_k), k\geq 1.
\]

By taking into account the \emph{geometric means} $\left( \rho_1,\rho_2,\ldots, \rho_k,\ldots\right)$ of the different ratios $\left(q_1, q_2,\ldots, q_k,\ldots \right) $, it is possible to write:
\begin{align*}
\rho_1& =  q_1, \, \text{mean of } q_1,  \, \text{namely } q_1,\\
\rho_2 &= \left(q_1 q_2\right) ^{1/2},\, \text{mean of } q_1  \, \text{and } q_2,\\
\rho_3 &=\left(q_1\, q_2\, q_3\right)^{1/3},\, \text{mean of } q_1, \,q_2 \, \text{and } q_3,\\
\ldots,\\
\rho_k &=\left(q_1\ldots q_k\right)^{1/k},\, \text{mean of } q_1, \ldots, q_k,\\
\ldots,
\end{align*} 
from where one draws, by squaring, cubing, ... :
\begin{align*}
 q_1&= \rho_1,\\
q_1 q_2 &=\rho_2^2,\\
q_1q_2 q_3 &=\rho_3^3,\\
\ldots,\\
q_1\ldots q_k &= \rho_k^k,\\
\ldots.
\end{align*} 

The terms of the sequence, expressed as functions of the different ratios, namely
\[
u_0,\, u_0q_1,\,u_0q_1q_2, \, \ldots,\, u_0q_1\ldots q_k,\, \ldots,
\]
can then be expressed as functions of the geometric means of the different ratios, namely
\[
u_0,\, u_0 \rho_1,\, u_0\rho_2^2,\, \ldots,\, u_0 \rho_k^k,\,\ldots, 
\]
the sum $s_k$ associated with this sequence, thus admitting an expression as a function of these means:
\[
s_k = u_0\left( 1+\rho_1 + \rho_2^2+\ldots+ \rho_k^k\right), \quad k\geq 1.
\]

\section{Why this kind of sequence with variable ratio} \label{sec:WhysuiteGeomVar}
We have introduced this kind of sequence ``with variable ratio'', as well as the associated sum, in order to conceive a viral spreading model whose \emph{internal structure} can be in conformity with the daily census process of the new contaminated.



In the census of the new contaminated by region and by day, that gives a space and time image of the spreading by contamination, there is no reason for the ratio of new contaminated between two consecutive days to be constant in time and therefore to be independent of the rank $k$. Indeed, the ratio between the new contaminated counted at day $k+1$ and the new contaminated counted at day $k$, assuredly depends on $k$, thus expressing that this ratio well varies day after day. 

That amounts to saying that such a census 
cannot be formalized by a simple geometric sequence, which is with constant ratio, whereas it can be well formalized by a geometric sequence ``with variable ratio'', whose ratio between two consecutive terms varies with the rank $k$. 

Moreover, given the complexity of contamination pheno\-mena, and particularly the stochastic aspects that highly contribute to this complexity, there is no reason for, day after day, the ratio of new contaminated between two consecutive days to increase or to decrease in a monotonic manner.
It is indeed very likely that this kind of variation can be obtained in \emph{mean}, in accordance with \emph{smoothing} this ratio:
hence, the geometric means of the different ratios, considered in our developments to go in this direction.

Such expectations, linked to our knowledge of complex phenomena, thus raise the points to be discussed, after confronting the proposed model to official data of the French Ministry of Health\footnote{https://www.data.gouv.fr/fr/datasets/donnees-relatives-a-lepidemie-de-covid-19-en-france-vue-densemble/}, which, from now on, will be named Ministry. 

That being said, it is now possible to formalize the daily census process of the new contaminated, and this, of course, in conformity with the geometric sequence ``with variable ratio''  (in the case of a spreading aggravation):
\begin{itemize}
\item at day $0$ (which corresponds to $k=0$ in the sequence), the term of rank $0$, $u_0$, represents the number of new contaminated individuals who have been counted this day;
\item at day $1$ (which corresponds to $k=1$ in the sequence), the term of rank $1$, $u_1$, greater than $u_0$, is the number of new contaminated individuals who have been counted this day, this number being written as $u_1=u_0 q_1$, thus expressing that at day $1$, there are $q_1$ times more new contaminated than the day before (day $0$);
\item at day $2$ (which corresponds to $k=2$ in the sequence), the term of rank $2$, $u_2$, greater than $u_1$, is the number of new contaminated individuals who have been recorded this day, this number being written as $u_2=u_1 q_2$, thus expressing that at day $2$, there are $q_2$ times more new contaminated than at day $1$;
\item at day $k$, the term of rank $k$, $u_k$, greater than $u_{k-1}$, can be written as $u_k=u_{k-1} q_k$, thus expressing that at day $k$, there are $q_k$ times more new contaminated than at day $k-1$.
\end{itemize}

The sum $s_k$ associated with the sequence, is written, as a function of the different ratios, according to
\[
s_k = u_0 \left( 1+ q_1 + q_1q_2+\ldots+ q_1\ldots q_k\right)\quad \text{for} \quad k\geq 1,
\]
or, as a function of the geometric means of the different ratios, according to
\[
s_k = u_0 \left( 1+ \rho_1 + \rho_2^2+\ldots+ \rho_k^k\right)\quad \text{for} \quad k\geq 1.
\]

This sum $s_k$, expressed by the one or the other of these two equations, represents the sum of the \emph{new} contaminated  who have been censed from day $0$ to day $k$ ($0$ and $k$ included). The total sum of the contaminated at day $k$, is then obtained by adding the \emph{former} contaminated censed before day $0$ (therefore until day $-1$).

\section{Impulse and step responses of non-integer integration operator and power-law}\label{App:Annexe1}

This section shows  that the power-law, $t^m$, can be defined as the step response of  a non-integer  order integrator, or as the solution  of a non-integer differential equation.

\subsection{Impulse response}\label{sub:ImpResp}
The unit impulse response, or response to the Dirac function, of the non-integer integration operator, of positive non-integer order, $m$, admits an expression of the form (\cite{Liouville1832,Erdelyi1962,Oustaloup2014}):
\[
y_{imp}(t) = \frac{t^{m-1}}{\Gamma(m)}u(t),
\]
$u(t)$ being the Heaviside function and $\Gamma(m)$ the gamma function defined by:
\[
\Gamma(m) = \int_0^{\infty} x^{m-1} e^{-x}dx.
\]

\subsection{Step response}
The unit step response, or response to the Heaviside function, of the non-integer integration operator, of positive non-integer order, $m$, is given by the integral of the unit impulse response, namely, for $t>0$:
\[
y_{step} (t) = \int_0^t y_{imp} (\theta)d\theta,
\]
or, given the expression of $y_{imp}(t)$:
\[
y_{step}(t) = \int_0^t \frac{\theta^{m-1}}{\Gamma(m)}d\theta = \frac{1}{\Gamma(m)} \int_0^t \theta^{m-1} d\theta,
\]
or else:
\[
y_{step}(t) = \frac{1}{\Gamma(m)}\left[ \frac{\theta^m}{m} \right]_0^t = \frac{1}{m\Gamma(m)} t^m = \frac{t^m }{\Gamma(m+1)} ,
\]
a result that enables concluding that the step response (unit or not) of the integration operator of non-integer order, $m>0$, well varies as the power-law $t^m$.

\subsection{A non-integer differential equation governing  the power-law $t^ m$}\label{App:Annexe2}
Given the step response so obtained, the power-law $t^m$ appears as a function  $f(t)$ such that:
\[
y_{step}  (t) = \frac{f(t)}{\Gamma(m+1)},
\]
or, in  Laplace transforms:
\[
Y_{step} (s)  = \frac{F(s)}{\Gamma(m+1)} \quad \text{with} \quad  F(s) = \mathcal{L} \left[f(t)  = t^m \right],
\]
a  symbolic equation that enables writing, 
knowing that 
\[
Y_{step} (s) = \frac{1}{s^m} U(s)\quad  \text{where} \quad U(s) =  \mathcal{L} \left[ u(t)\right]:
\]
\[
 \frac{F(s)}{\Gamma(m+1)}  =  \frac{U(s)}{s^m}, 
\]
namely:
\[
s^m F(s) = \Gamma(m+1) U(s),
\]
from where one draws the concrete equation:
\[
\left( \frac{d}{dt}\right)^m f(t)  = \Gamma(m+1) u(t),
\]
a relation that expresses a linear differential equation of non-integer order, $m$, whose solution is nothing but the power-law $f(t) = t^m$.

\section{Proposition and validation of a time model of viral sprea\-ding:  the FPM model} \label{sec:TimeModelSpreading}
\begin{figure}[h]
\hspace{-0.6cm}
 \psfragfig[width=1.15\columnwidth]{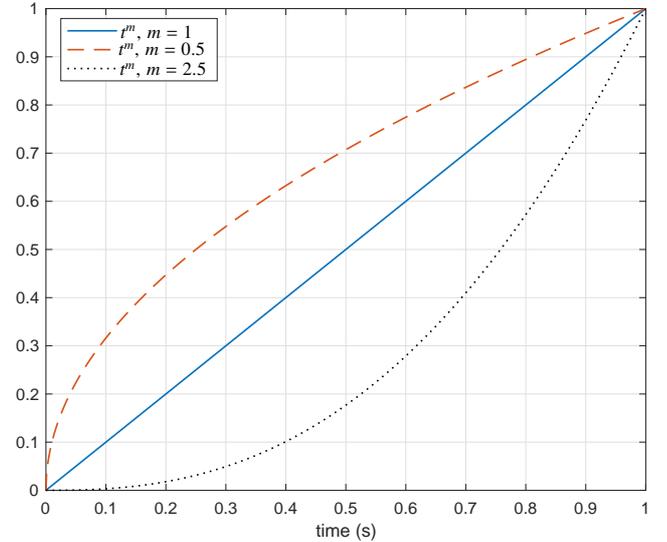}{
  \psfrag{aaaaaaa}[c][][0.7]{\hspace{0.2cm}$t^m, \, m =1$}
   \psfrag{bbbbbbb}[c][][0.7]{\hspace{0.4cm}$t^m, \, m =0.5$}
    \psfrag{ccccccccccc}[c][][0.7]{$t^m, \, m =2.5$}
    }
\caption{Step response in $t^m$, whose non-integer power, $m$, determines the curvature, namely a concavity for $m<1$ and a convexity for $m>1$: for $m<<1$, $t^m=e^{\ln t^m}=e^{m \ln t}$, is reduced to $1+m \ln t$, whose variation in $\ln t$ is representative of a long memory phenomenon}
\label{fig:StepR}
 \end{figure}

Given that the non-integer differentiation or integration operator governs diffusion phenomena (thermal for example), one can legitimately suppose that this operator of non-integer order, $-m$ in differentiation or $m$ in integration, with $m>0$ , and of step response in $t^m$ (section \ref{App:Annexe1} and figure \ref{fig:StepR}), also governs spreading phenomena by contamination. It is true that these phenomena, as the diffusion phenomena, fall under \emph{complex phenomena}, for which the non-integer operator or its dynamics in power-law turn out to be particularly appropriated mo\-deling tools, notably by the \emph{parametric compactness} they offer in comparison with complexity.
A newsworthy illustration that widens this context, incontestably lies in the field of huge networks and particularly of Internet network: Internet is indeed a paradigm of complex network, whose ``degree distribution", which is inherent to the different average numbers of connections per node and which thus evokes a ``diversity degree'' (\cite{Oustaloup2014}), follows a non-integer power law such as $N(k) = k^{-\gamma}$, $N(k)$ representing the number $N$ of nodes endowed with $k$ connections, and $\gamma$ being a positive non-integer comprised between 2 and 3 (\cite{Santamaria2017} and \cite{Mendes2009}).
Other applicative fields in which the power-law dynamics occurs are presented in \ref{App:AnnexePower-law}.

To properly verify such an hypothesis of non-integer exponent, {{besides \emph{argued} by a \emph{regular distribution of internal dynamics (sections \ref{subsec:ComplexityInternalStructure} and \ref{sec:SystemicApproach})},}} it suffices to directly check the conformity of a variation in $t^m$ with the official data of Ministry, notably according to the comparative analysis presented further and which well proves this conformity (section \ref{sec:Validation_tm_COVID19}).

But it is possible to go beyond this direct verification, by determining the different ratios and the geometric means of the different ratios, inherent to the hypothesis of a variation in $t^m$, and this, to see if, themselves, are well in conformity with the official data of the Ministry. This is a way to attempt to validate the internal structure proposed for the viral spreading model. 

To that effect, a first step consists in imposing, to the sum of the new contaminated, $s_k$, a time variation in $t^m$ (up to a multiplying constant and an additive constant), which amounts to simply writing, by introducing a positive constant, $c$:
\[
s_k = u_0\left( 1+ct^m \right),
\]
namely, for $t=kh$ where $h$ is the sampling period:
\[
s_k = u_0\left(1+c(kh)^m \right),
\]
therefore, given the expressions of $s_k$ such as already determined:
\[
 u_0 \left( 1+ q_1 + q_1q_2+\ldots+ q_1\ldots q_k\right) = u_0\left(1+c(kh)^m \right), \,k\geq 1
\]
and 
\[
  u_0 \left( 1+ \rho_1 + \rho_2^2+\ldots+ \rho_k^k\right)= u_0\left(1+c(kh)^m \right), \,k\geq 1,
\]
or, after simplifying by $u_0$ and $1$:
\begin{equation}\label{eq:first}
q_1 + q_1q_2+\ldots+ q_1\ldots q_k = c(kh)^m \quad \text{with} \quad k\geq 1
\end{equation}
and 
\begin{equation}\label{eq:second}
  \rho_1 + \rho_2^2+\ldots+ \rho_k^k=c(kh)^m  \quad \text{with} \quad k\geq 1.
\end{equation}

From the two equations so obtained, a second step consists in determining the different ratios, $q_1$, $q_2$, $\ldots$, $q_k$, as well as the geometric means of the different ratios, $\rho_1$, $\rho_2$, $\ldots$, $\rho_k$, for a sampling period of one day, namely $h=1$, and for the different values of $k$, namely $1$, $2$, $3$, $\ldots$

\subsection{Determination of the different ratios} \label{subsec:RatioDetermination}

For $h=1$, the first equation \eqref{eq:first} becomes: 
\[
q_1 + q_1q_2+\ldots+ q_1\ldots q_k = ck^m \quad \forall k\geq 1.
\]

For $k=1$, a term by term identification immediately gives:
\[
q_1 = c1^m = c.
\]

For $k\geq 2$, let us rewrite the equation in conformity with:
\begin{equation}\label{eq:first1}
q_1 + q_1q_2+\ldots+q_1\ldots q_{k-1} + q_1\ldots q_k = ck^m,
\end{equation}
then let us replace $k$ by $k-1$:
\begin{equation}\label{eq:first2}
q_1 + q_1q_2+\ldots+q_1\ldots q_{k-2} + q_1\ldots q_{k-1} = c\left( k-1\right)^m.
\end{equation}

The difference of equations \eqref{eq:first1} and \eqref{eq:first2} leads to:
\begin{equation}\label{eq:first3}
 q_1\ldots q_{k} = c\left( k^m - \left( k-1\right)^m\right),
\end{equation}
or, by replacing $k$ by $k-1$:
\begin{equation}\label{eq:first4}
 q_1\ldots q_{k-1} = c\left( \left( k-1\right)^m - \left( k-2\right)^m\right),
\end{equation}
the ratio of equations \eqref{eq:first3} and \eqref{eq:first4} then leading to:
\begin{equation}\label{eq:first5}
q_k = \frac{k^m - \left( k-1\right)^m}{\left( k-1\right)^m-\left( k-2\right)^m}\quad \forall k\geq 2,
\end{equation}
a result that expresses the general term of the sequence $\left ( q_k\right)_{k\geq 2}$.

For $k>> 1$, it is convenient to write $q_k$ under the form:
\[
\begin{split}
q_k &= \frac{k^m - \left[ k\left( 1 - \frac{1}{k}\right)\right]^m}{\left[ k\left( 1 - \frac{1}{k}\right)\right]^m-\left[ k\left( 1 - \frac{2}{k}\right)\right]^m}\\
&=  \frac{k^m - k^m \left( 1 - \frac{1}{k}\right)^m}{k^m \left( 1 - \frac{1}{k}\right)^m-k^m \left( 1 - \frac{2}{k}\right)^m},
\end{split}
\]
or, by simplifying by $k^m$:
\[
q_k =  \frac{1 - \left( 1 - \frac{1}{k}\right)^m}{ \left( 1 - \frac{1}{k}\right)^m- \left( 1 - \frac{2}{k}\right)^m},
\]
namely, for $k$ sufficiently high:
\[
q_k \sim  \frac{1- \left( 1 - \frac{m}{k}\right)}{ \left( 1 - \frac{m}{k}\right)- \left( 1 - \frac{2m}{k}\right)} = \frac{m/k}{m/k},
\]
therefore, finally:
\[
q_k \sim 1\quad \forall k>>1,
\]
a result, not only remarkable of simplicity, but also very interesting with regard to the official data of the Ministry.

\begin{figure}[h]
\hspace{-0.6cm}
 \psfragfig[width=1.15\columnwidth]{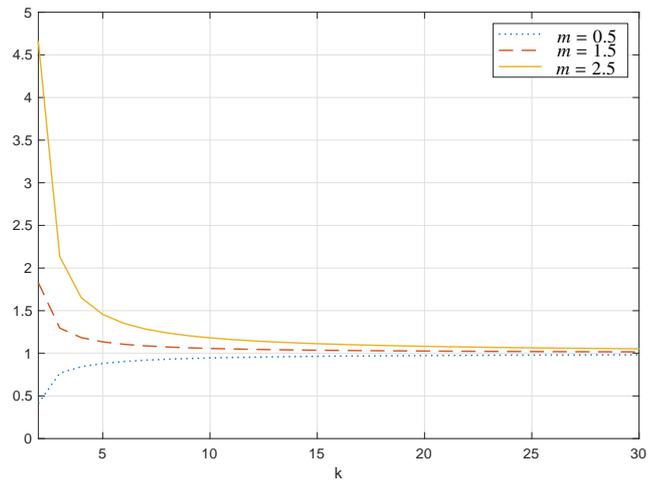}{
  \psfrag{aaaaaaaaaa}[c][][0.7]{\hspace{0.2cm}$m =0.5$}
   \psfrag{bbbbbbbbbb}[c][][0.7]{\hspace{0.2cm}$ m =1.5$}
    \psfrag{cccccccccc}[c][][0.7]{\hspace{0.3cm}$m =2.5$}
    }
\caption{Variations of $q_k$ versus $k$ for three values of $m$}
\label{fig:Var_qk_m}
\end{figure}

Figure \ref{fig:Var_qk_m} illustrates the variations of $q_k$ given by \eqref{eq:first5} for different values of $m$ ($m=0.5,\, 1.5$ and $2.5$).

\subsection{Determination of the geometric means of the different ratios} \label{subsec:MeanRatioDetermination}

For $h=1$, the second equation \eqref{eq:second} becomes:
\[
  \rho_1 + \rho_2^2+\ldots+ \rho_k^k=ck^m  \quad \forall k\geq 1.
\]

For $k\geq1$, let us rewrite this equation under the form:
\begin{equation}\label{eq:second1}
  \rho_1 + \rho_2^2+\ldots+ \rho_{k-1}^{k-1}+\rho_k^k=ck^m,
\end{equation}
then let us replace $k$ by $k-1$:
\begin{equation}\label{eq:second2}
  \rho_1 + \rho_2^2+\ldots+ \rho_{k-2}^{k-2}+\rho_{k-1}^{k-1}=c(k-1)^m.
\end{equation}

The difference of equations \eqref{eq:second1} and \eqref{eq:second2} leads to:
\begin{equation*}
 \rho_k^k=c\left(k^m- (k-1)^m\right),
\end{equation*}
from where one draws:
\begin{equation}\label{eq:rhok}
 \rho_k=\left( c^{\frac{1}{k}}\right)\left(k^m- (k-1)^m\right)^{\frac{1}{k}} \quad \forall k\geq 1,
\end{equation}
a result that expresses the general term of the sequence $\left ( \rho_k\right)_{k\geq 1}$.

For $k>>1$, it is appropriated to write:
\[
\begin{split}
k^m- (k-1)^m&= k^m- \left[k\left (1-\frac{1}{k}\right)\right]^m\\
& = k^m- k^m\left(1-\frac{1}{k}\right)^m\\
&= k^m \left[ 1-\left( 1 - \frac{1}{k}\right) ^m \right],
\end{split}
\]
namely, for $k$ sufficiently high:
\[
k^m- (k-1)^m \sim k^m\left[1 - \left(1- \frac{m}{k}\right) \right] = k^m \frac{m}{k}.
\]

$\rho_k$ is then written:
\[
\rho_k = \left(c^{\frac{1}{k}} \right) \left(k^m - (k-1)^m\right)^{\frac{1}{k}} \sim \left(c^{\frac{1}{k}} \right) \left(k^m \frac{m}{k}\right)^{\frac{1}{k}} ,
\]
or:
\[
\begin{split}
\rho_k &\sim \left(c^{\frac{1}{k}} \right)k^\frac{m}{k} \left(\frac{m}{k}\right)^{\frac{1}{k}} \\
& =  \left(c^{\frac{1}{k}} \right)\frac{k^\frac{m}{k}}{k^\frac{1}{k}} m^\frac{1}{k}\\
&=  \left(cm \right)^{\frac{1}{k}} k ^\frac{m-1}{k},
\end{split}
\]
or even, knowing that
\[
k^{\frac{m-1}{k}} = e^{\ln\left( k ^{\frac{m-1}{k}}\right)} = e^{\frac{m-1}{k}\ln k } = e^{(m-1) \frac{\ln k }{k}} :
\]
\[
\rho_k \sim\left(cm \right)^{\frac{1}{k}} e^{(m-1) \frac{\ln k }{k}},
\]
or else, given that for $k$ sufficiently high, 
\[
\begin{split}
\frac{1}{k} \sim 0 \quad \text{and} \quad   \frac{\ln k }{k}\sim 0:\\
\rho_k \sim 1 \quad \forall k>>1,
\end{split}
\]
a result, not only remarkable of simplicity, but also very interesting with respect to the official data of the Ministry. 

\begin{figure}[h]
\hspace{-0.6cm}
 \psfragfig[width=1.15\columnwidth]{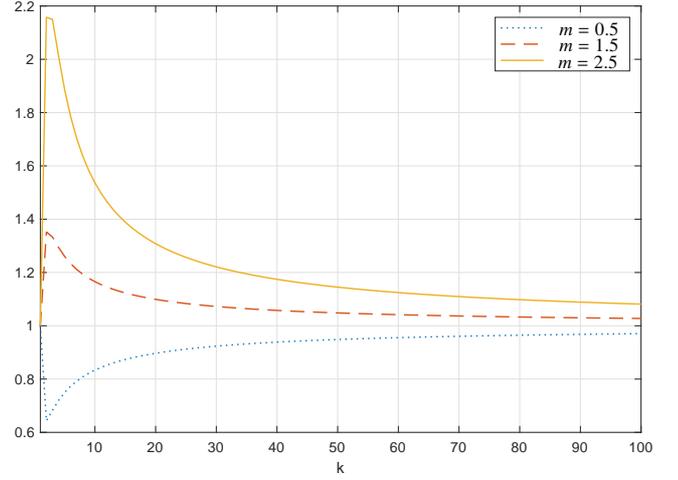}{
  \psfrag{aaaaaaaaaa}[c][][0.7]{\hspace{0.2cm}$m =0.5$}
   \psfrag{bbbbbbbbbb}[c][][0.7]{\hspace{0.2cm}$ m =1.5$}
    \psfrag{cccccccccc}[c][][0.7]{\hspace{0.3cm}$m =2.5$}
    }
\caption{Variations of $\rho_k$ versus $k$ for $c=1$ and three values of $m$}
\label{fig:Var_rhok_m}
\end{figure}

Figure \ref{fig:Var_rhok_m} presents, for $c=1$, the variations of $\rho_k$ given  by \eqref{eq:rhok} for different values of $m$ ($m=0.5,\, 1.5$ and $2.5$).

\subsection{Validation of the internal structure of the model} \label{subsec:Validation_tm}
\begin{figure}[th]
\hspace{-0.6cm}
 \psfragfig[width=1.15\columnwidth]{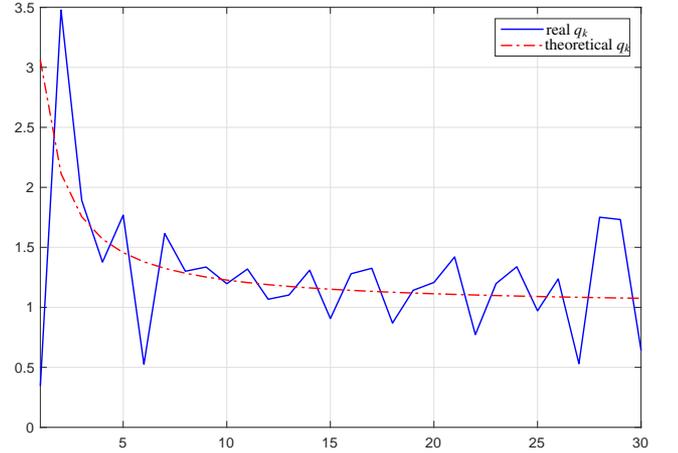}{
  \psfrag{bbbbbbbbbb}[c][][0.6]{\hspace{0.1cm} theoretical $q_k$}
   \psfrag{aaaaaaaaaa}[c][][0.6]{\hspace{-0.8cm} real $q_k$}
    }
\caption{Comparative illustration of the theoretical and real $q_k$ for $m=3.2523$: the theoretical $q_k$ smooth the real $q_k$ ;  
$\textrm{mean}(E_r) =0.0078$}
\label{fig:4}
\end{figure}

\begin{figure}[h]
\hspace{-0.6cm}
 \psfragfig[width=1.15\columnwidth]{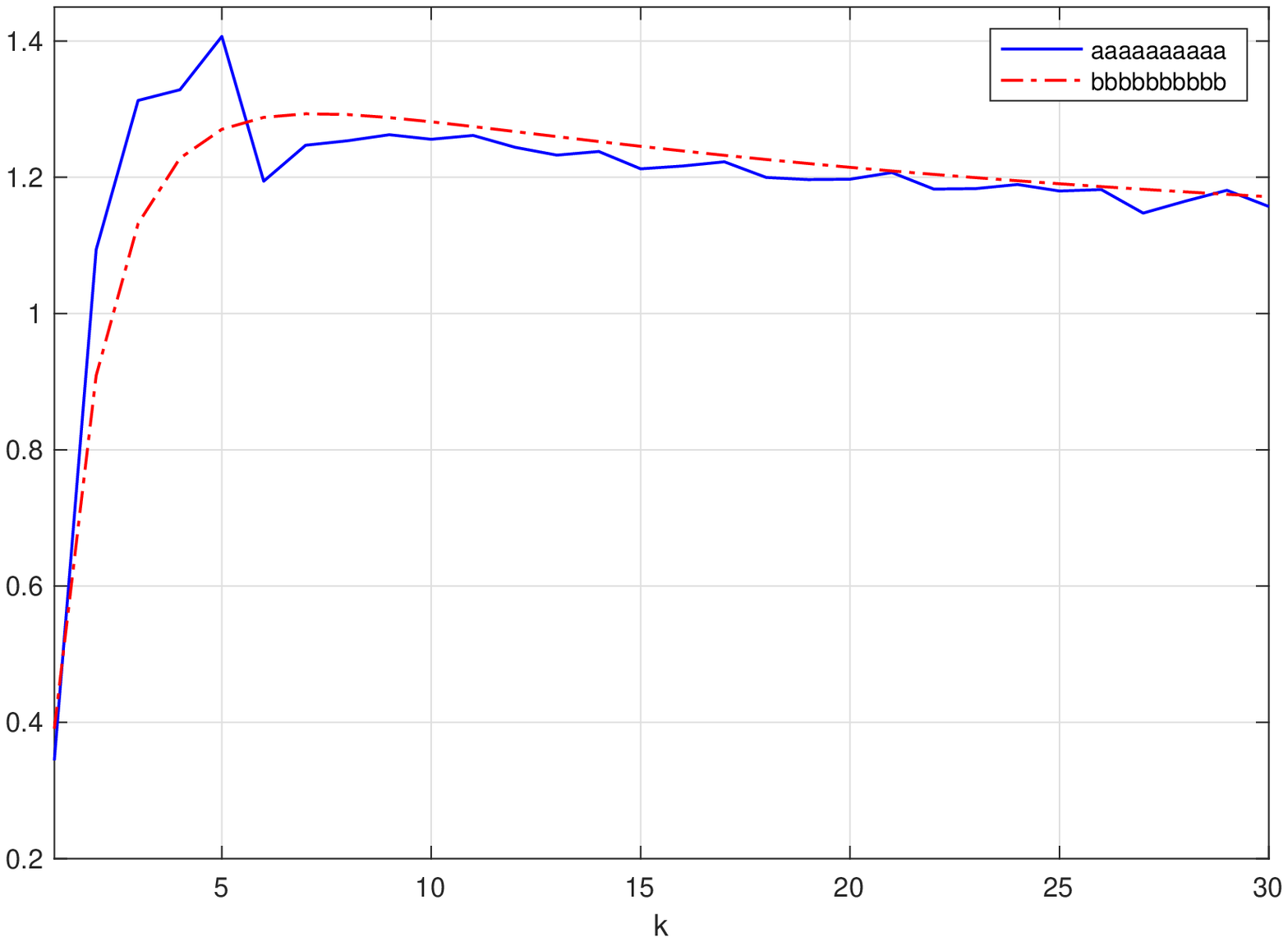}{
  \psfrag{bbbbbbbbbb}[c][][0.6]{\hspace{0.1cm} theoretical $\rho_k$}
   \psfrag{aaaaaaaaaa}[c][][0.6]{\hspace{-0.8cm} real  $\rho_k$}
    }
\caption{Comparative illustration of the theoretical and real $\rho_k$ for $m=3.2523$ and $c=0.01497$: 
the theoretical $\rho_k$ smooth the real $\rho_k$ ; $\textrm{mean}(E_r) =-4.1497\, 10^{-4}$}
\label{fig:5}
\end{figure}

\begin{figure}[th]
\hspace{-0.6cm}
 \psfragfig[width=1.15\columnwidth]{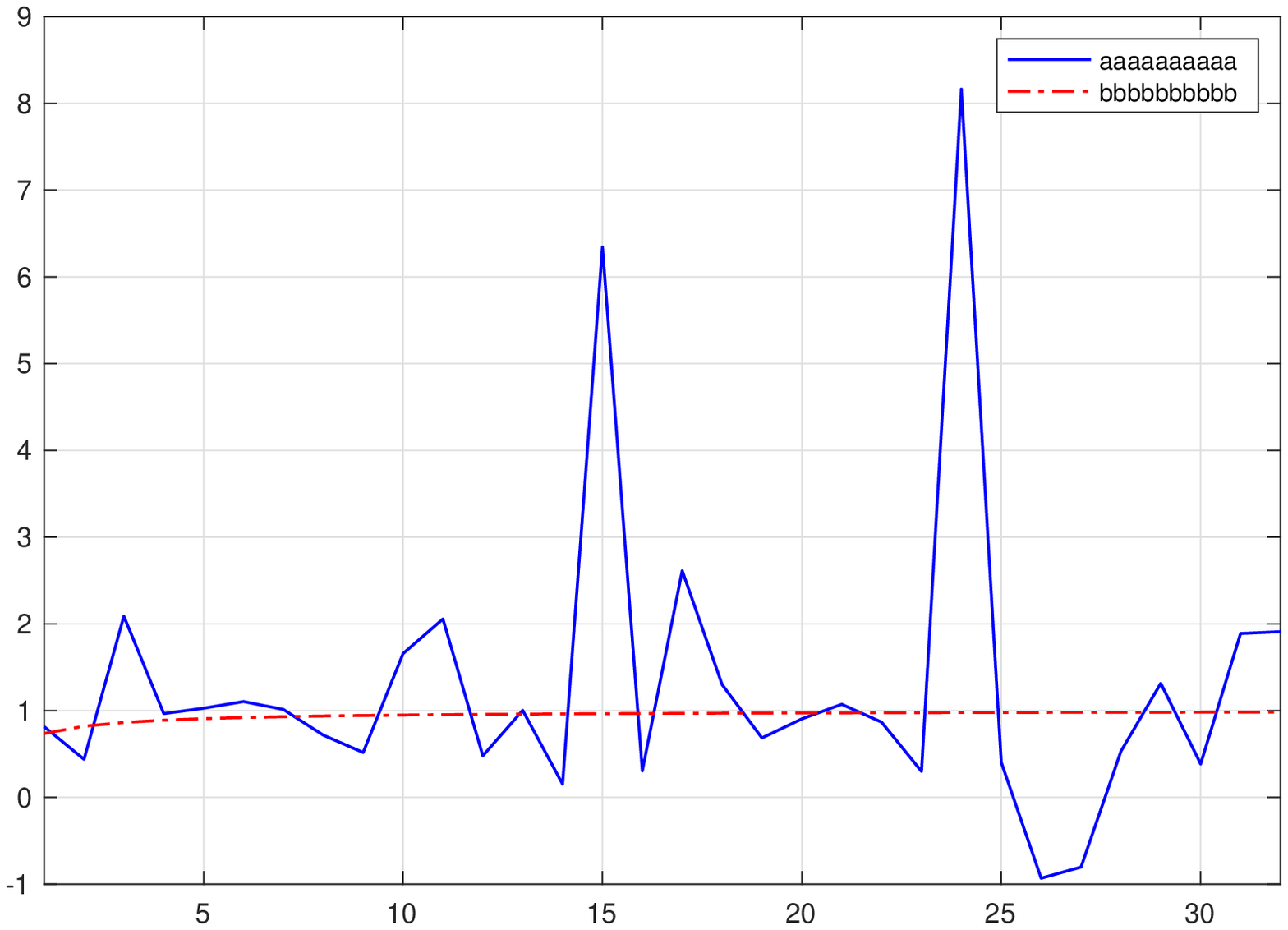}{
  \psfrag{bbbbbbbbbb}[c][][0.6]{\hspace{0.1cm} theoretical $q_k$}
   \psfrag{aaaaaaaaaa}[c][][0.6]{\hspace{-0.8cm} real  $q_k$}
    }
\caption{Comparative illustration of the theoretical and real $q_k$ for $m=0.4221$: the theoretical $q_k$ smooth the real $q_k$ ; 
$\textrm{mean}(E_r) =0.3591$ because of the aberrations due to incorrect data }
\label{fig:6}
\end{figure}

The random fluctuations of the ratios, inherent to the reality complexity, are, properly, interpretable as fluctuations that are superimposed to a \emph{smoothing} whose graphical representation is a curve representative, no more of the variations of these ratios, but of the variation mean.

So, the internal structure of the model in $t^m$, such as elaborated by disregarding the unpredictable (these random fluctuations), can only correspond to the smoothing as defined. 

That is to say that the ratios $q_k$ must be in conformity with the smoothing of the ratios stemming from the official data of Ministry. Thus, the validation of the proposed internal structure, can be carried out by the verification of this conformity. 

\begin{figure}[h]
\hspace{-0.6cm}
 \psfragfig*[width=1.15\columnwidth]{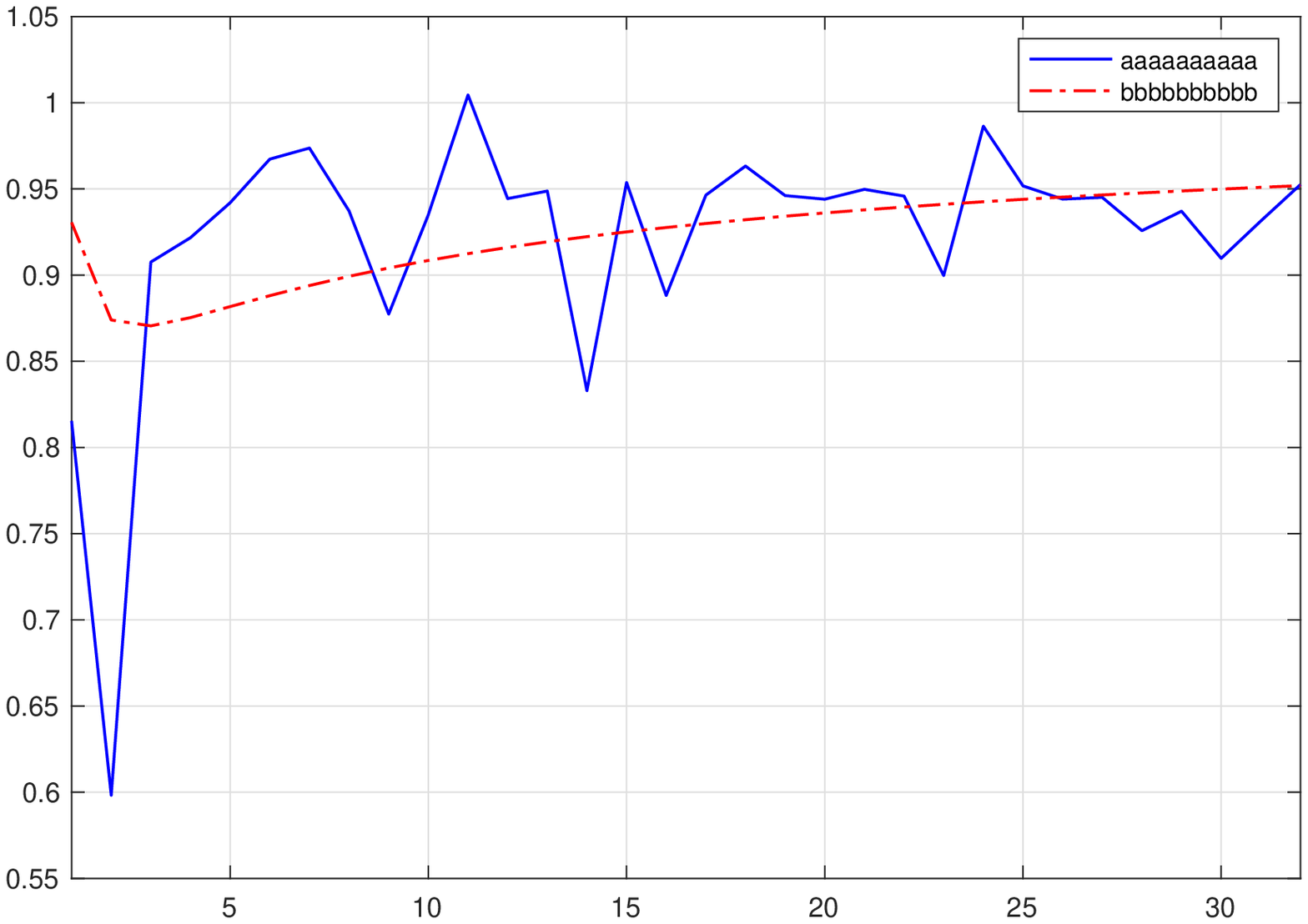}{
  \psfrag{bbbbbbbbbb}[c][][0.6]{\hspace{0.1cm} theoretical $\rho_k$}
   \psfrag{aaaaaaaaaa}[c][][0.6]{\hspace{-0.5cm} real  $\rho_k$}
    }
\caption{Comparative illustration of the theoretical and real $\rho_k$ for $m=0.4221$ and $c=3.7202$: 
the theoretical $\rho_k$ smooth the real $\rho_k$;  $\textrm{mean}(E_r) =7.9843\, 10^{-5}$}
\label{fig:7}
\end{figure}
But this verification can be reinforced by benefiting from the study of the geometric means $\rho_k$ of the different ratios $q_k$, notably by comparing the geometric means obtained, theoretically, with the internal structure and, really, with the official data of Ministry. 

If $a_k$ denotes the total number of contaminated counted at day $k$ by the Ministry, the real $q_k$ and $\rho_k$ $\forall k\geq 1$, stemming from the official data, are computed in conformity with:
\[
q_k = \frac{a_{k+1}-a_{k}}{a_{k}-a_{k-1}}
\]
and 
\[
\rho_k = \left ( q_1\ldots. q_k \right)^{1/k},
\]
where:
$$\begin{aligned}
q_1 \ldots q_k &= \frac{a_2-a_1}{a_1 - a_0} \frac{a_3-a_2}{a_2 - a_1} \ldots \frac{a_{k}-a_{k-1}}{a_{k-1} - a_{k-2}}\frac{a_{k+1}-a_{k}}{a_{k} - a_{k-1}} \\
&=\frac{a_{k+1}-a_{k}}{a_{1} - a_{0}}.
\end{aligned}$$

To evaluate the quality of the modeling, the maximum of the modulus of the absolute or relative errors, or the mean of the absolute or relative errors can be computed, namely ($a_k$ being the real data and $b_k$ the theoretical values stemming from the modeling):
\[
\max\left (|E_a|\right) = \max \{| a_k - b_k|,\, 1\leq k\leq n\} 
\]
or
\[
\textrm{ mean}\left(E_a\right) = \frac{1}{n} \sum_{k=1}^n \left (a_k-b_k\right)
\]
or
\[
\max\left( |E_r |\right) = \max \left\{\left | \frac{a_k - b_k}{b_k}\right |,\, 1\leq k\leq n\right\} \, 
\]
or
\[
\textrm{mean}\left(E_r\right) = \frac{1}{n} \sum_{k=1}^n \left( \frac{a_k-b_k}{b_k}\right ) .
\]

To lead the verification of the conformity of $q_k$ and $\rho_k$ with the smoothing of the corresponding quantities stemming from the official data, we have taken two values of $m$ stemming from the identification, by the model in $t^m$, of the COVID-19 spreading in France (section \ref{sec:Validation_tm_COVID19}): 
the first value of $m$, greater than 1, namely $m=3.252$, corresponds to the convexity that period 1 presents, of 32 days, from March 1 to April 1 $2020$ ; the se\-cond value of $m$, less than 1, namely $m=0.422$, corresponds to the concavity, that period 2 presents, of $34$ days, from April $2$ to May $5$. Concerning the adjustment parameter, $c$, that takes part in the general expression of $\rho_k$, its determination according to the relation $c=B/u_0$ drawn from \eqref{eq:A_B}, is not pertinent in the sense that, if $B$ is known (section \ref{sec:Validation_tm_COVID19}), $u_0$ is known only through a noisy data: so, $c$ is determined by optimization, so that the arithmetic mean of the absolute gaps, $mean(E_a)$, should be theoretically null. Thus, we get $c=0.01497$  
for period 1 (figure \ref{fig:5}) and  $c=3.7202$ 
for period 2 (figure \ref{fig:7}).  

For the first values of $k$, figures \ref{fig:5} and \ref{fig:7} well show that the geometric mean of noisy data is not a very significative mean. 
Note that the data can present some aberrations, particularly those collected at the weekend. 

From the whole of the curves represented by figures \ref{fig:4}  to \ref{fig:7}, it turns out that the conformity to be verified is truly effective, thus validating the internal structure of the model in $t^m$, and consequently, the model itself (more precisely the model in $t^m$ up to constants) as the internal structure is inherent to this model.


\subsection{Validation of the self-filtering action of the internal structure of  the model} \label{subsec:ValidationAction_tm}
\begin{figure}[h]
\hspace{-0.6cm}
 \psfragfig[width=1.15\columnwidth]{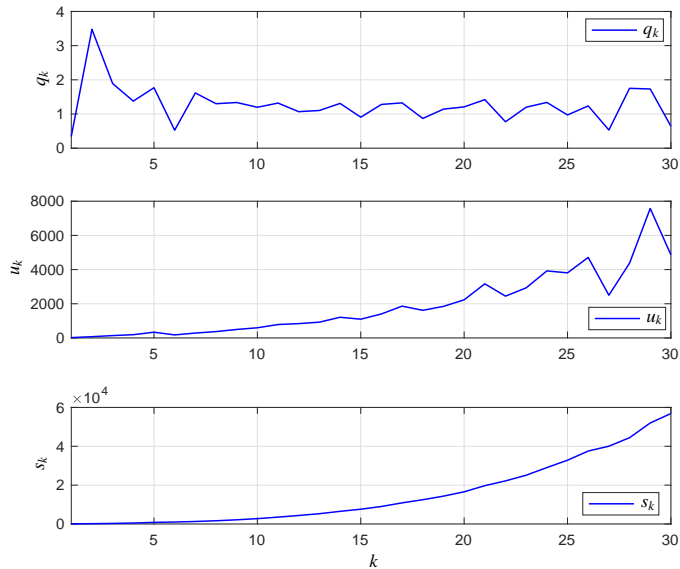}{
  \psfrag{qk}[c][][0.7]{ $q_k$}
  \psfrag{uk}[c][][0.7]{ $u_k$}
    \psfrag{sk}[c][][0.7]{ $s_k$}
      \psfrag{k}[c][][0.7]{ $k$}
    }
\caption{Illustration of the self-filtering action of the internal structure, from the real data, $a_k$, $k\geq 0$, corresponding to period 1. For $k\geq 1$: $q_k = (a_{k+1}-a_k)/(a_k-a_{k-1})$; $u_k = u_0 q_1 \ldots q_k = a_{k+1} - a_k $ with  $u_0 = a_1  - a_0$;  $s_k = u_0 + u_1+ \ldots + u_k$}
\label{fig:qk_uk_sk}
\end{figure}

The validation presented in the previous section \ref{subsec:Validation_tm} is enriched by the validation of the \emph{self-filtering action} of the internal structure, such as described at the end of section \ref{subsec:contributions} on the specific contributions of the article. This action, that makes the model in $t^m$ not noisy even if it is internally noisy, indeed turns out to be graphically validated by figure \ref{fig:qk_uk_sk}, obtained with the real data of period 1 which corresponds to a convex growth.

The curves of this figure well convey two successive reductions of the random fluctuations of the ratios $q_k$ between new contaminated between two consecutive days: it is, on one hand, a reduction between the $q_k$ and  the $u_k$ (effect of a first filtering) and, on the other hand, a reduction between the $u_k$ and the $s_k$ (effect of a second filtering), the $u_k$ denoting the new contaminated day after day, and the $s_k$ denoting the daily sums of these new contaminated. Actually, through all these curves, figure \ref{fig:qk_uk_sk}  speaks by itself: indeed, this figure, which well illustrates two successive improvements of the signal-to-noise ratio, is sufficiently convincing for not being more commented.

The only comments are of another nature as they turn on the origin of the first filtering effect. At section \ref{subsec:contributions}, we have attributed this origin to the form of the sequence terms, therefore to the products of the different ratios,  $q_1\ldots q_k$, given the considered  geometric sequence ``with variable ratio''. But if we had considered an arithmetic sequence ``with variable difference'', the  origin of  the first filtering would have been attributable to the sums of the different differences, $r_1+\dots + r_k$, stemming  from the general term of the sequence, namely $u_k = u_0 + r_1 + \ldots + r_k$, $\forall k \geq 1$. These sums of differences, that present a discrete integral action, and which are thus likely to filter themselves the random fluctuations of the differences, offer an explanation, as or even more perceptible as the one founded on the geometric  sequence ``with variable ratio''. Note that this sequence was preferred knowing that the  $q_k$ are always positive, contrary to the $r_k$. 

Thus, more generally, for an internal structure based on a geometric or arithmetic sequence, its self-filtering  property results from the form itself of the structure, which (in reality) globally eliminates the random fluctuations of its elements, thanks to the change from the \emph{local} defined  by the $q_k$ or $r_k$, to the \emph{global} defined by the $s_k$, namely the sums  of the $u_k$. Such a property is admittedly likely to explain the compatibility between the \emph{internal complexity} that the internal structure  and its stochastic behavior present, and the \emph{global simplicity} that  the  deterministic  model offers through its power-law, $t^m$. But this  simplicity form is not usual, in the sense that $t^m$ presents, not only a mathematical structure remarkable of simplicity, but also a great richness by expressing very simply the complexity of reality, as if  $t^m$, through its non-integer power $m$, \emph{implicitly} had a complexity of another kind. Actually, the change to the global is accompanied by a transformation of the complexity nature: indeed, the internal complexity globally comes down to a non-integer power. It is true that the non-integer character of the power expresses or represents the complexity of systems and complex phenomena. That is to say that  the non-integer power of a power-law is a marker of complexity. Given this context, it is  convenient to recall that the  power-law $t^m$ is none other than  the solution of a differential equation (section \ref{App:Annexe1}) whose non-integer order, $m$, expresses the  character of infinite dimension (therefore the complexity) of the non-integer differentiation or integration operator, an operator which is itself  (in the operational domain) a  power-law, but of the operational variable, $s$.

\subsection{A FPM model that takes into account the totality of the  contaminated: associated differential equation} \label{subsec:ModelContamine}

After  validating the internal structure inherent  to the model  in $t^m$, thus validating the model itself, it only remains to test the precision of this model, therefore its representativity, with the official data provided by the Ministry on the COVID-19 spreading. Knowing that these data relate to all the (new and former) contaminated, it is convenient to complete the model so that it represents the whole set of the contaminated, that is to say the total sum of contaminated at day $k$, $S_k$, resulting from the sum of new contaminated recorded from day $0$ to day $k$, $s_k$, and the sum of the former contaminated recorded before day $0$, noted here $C_f$, namely:
\[
S_K=s_k+C_f,
\]
or, given the expression of $s_k$ for $t=kh$ and $h=1$:
\[
S_k = u_0\left (1+ck^m\right) +C_f,
\]
therefore, finally:
\begin{equation}\label{eq:A_B1}
S_k = A+Bk^m,
\end{equation}
with: 
\begin{equation}\label{eq:A_B}
A= u_0+C_f\quad \text{and}\quad B=u_0c.
\end{equation}

\label{ED_FPM}
{{
Given this model, it is easy to establish the differential equation that governs it. By denoting the model by $C(t)$, namely
\[
C(t) = A+Bt^m,
\]
one can use this equation to deduce the time, $t$:
\[
t = \left ( \frac{C(t) - A}{B} \right)^{\displaystyle\frac{1}{m}}.
\]
The differentiation of $C(t)$ is immediate, namely:
\[
\frac{dC(t)}{dt} = Bm t^{m-1},
\]
or, by replacing the time, $t$, by its expression such as determined:
\[
\frac{dC(t)}{dt} = Bm\left ( \frac{C(t) - A}{B} \right)^{\displaystyle\frac{m-1}{m}},
\]
a result that expresses a \emph{nonlinear differential equation with non-integer power}.
}}

\section{Time identification of the COVID-19 spreading by application of the FPM model to the official data on the contaminated} \label{sec:Validation_tm_COVID19}
This section is dedicated to the time identification, by a non-integer model, of the COVID-19 spreading in France, based on data between March 1 and October 1 $2020$, namely on $7$ months or $215$ days.

{{More precisely, the identification method that uses the least squares based on a nonlinear optimisation algorithm,}}\label{rev1:Q3} is founded on an \emph{identification model} whose structure is given, namely
\begin{equation}\label{eq:ABtm}
A+Bt^m,
\end{equation}
and whose three parameters, to be determined in an optimal manner and without constraint, are respectively the additive constant, $A$, the multiplicative constant, $B$, and the non-integer exponent, $m$, which makes the specificity of the model. 

\begin{figure}[h]
\hspace{-0.6cm}
 \psfragfig[width=1.15\columnwidth]{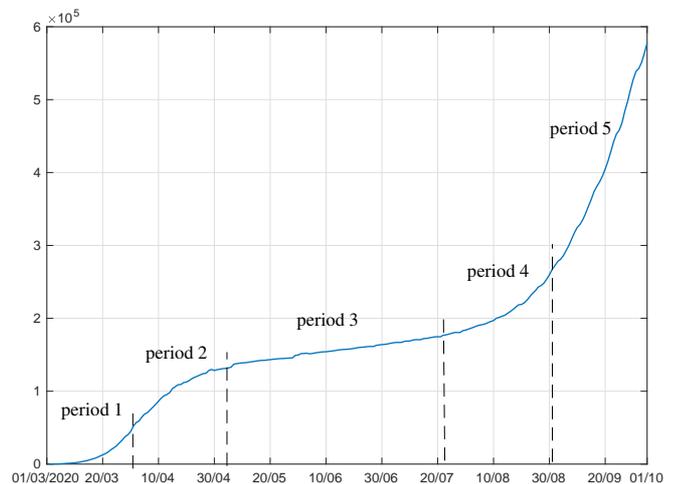}{
  \psfrag{period1}[c][][0.7]{\hspace{0.1cm} period 1}
  \psfrag{period2}[c][][0.7]{\hspace{-0.2cm} period 2}
    \psfrag{period3}[c][][0.7]{\hspace{-0.1cm} period 3}
      \psfrag{period4}[c][][0.7]{\hspace{-0.1cm} period 4}
        \psfrag{period5}[c][][0.7]{\hspace{-0.2cm} period 5}
    }
\caption{Evolution of the confirmed cases $a_i$ between March 1 and October 1 2020, on 215 days}
\label{fig:8}
\end{figure}

The data of the Ministry on the total number $a_i$ of the confirmed cases at day $i$, begin at March 1 2020 and are gathered in figure \ref{fig:8}. This figure presents several time periods for which the curve form is different: the curve seems convex for the first 32 days, then concave for the following 33 days (from day 33 to day 66), then quasi-linear from day 67 to day 134, and finally convex. 
If the curve is convex, $m$ is greater than $1$: the daily number of confirmed new cases increases and the epidemic progresses. If the points $(k,a_{d+k})$ are aligned, then $m=1$.
If the curvature is concave, then $0<m<1$: the daily number of confirmed new cases decreases and the epidemic regresses. If the daily number of confirmed new cases is null, $m=0$ and the epidemic stops. 

For the period between days $d+1$ and $d+n$, where $d$ denotes the day preceding this period and where $n$ is the number of days of this period, a time modeling is sought under the form \eqref{eq:ABtm}. That is to say that $a_{d+k}$ is approximated by an expression of the form $A+Bk^m$ where $k$ varies from 1 to $n$. 

To that effect, the least squares are used to determine the three parameters $A>0$, $B>0$ and $m>0$ that minimize the gap square sum. Given that the parameters $A$, $B$ and $m$ are calculated at best by optimization, on each period, the arithmetic mean of the absolute errors, $mean(E_a)$ is theoretically null. More precisely, if we operate with 20 significative digits, we find, for example, $mean(E_a) = -0.4\, 10^{-13}$ for period 1, and $mean(E_a) = -0.10\, 10^{-12}$ for period 2. 

For period 1, between March 1 and April 1 2020, namely 32 days, where $d=0$, $n=32$ and $1\leq k\leq 32$, the identification procedure leads to:
\[
A = 434.8972; \, B = 0.7148;\,  m= 3.2523.
\]

For period 2, between April 2 and May 5 2020, namely 34 days, where $d=32$, $n=34$ and $1\leq k\leq 34$, we  get:
\[
A = 30809.42; \, B = 23934.54;\,  m= 0.4221.
\]

For period 3, between May 6 and July 22 2020, namely 78 days, where $d=66$, $n=78$ and $1\leq k\leq 78$, we  get:
\[
A = 136349.46; \, B = 562.9646;\,  m= 0.9761.
\]

For period 4, between July 23 and September11 2020, namely 51 days, where $d=144$, $n=51$ and $1\leq k\leq 51$, we  get:
\[
A = 183943.7; \, B = 15.6681;\,  m= 2.3730.
\]

For period 5, between September 12 and October 1 2020, namely 20 days, where $d=195$, $n=20$ and $1\leq k\leq 20$, we get:
\[
A = 363041.1; \, B = 7300.427;\,  m= 1.13179.
\]

The following figures, from \ref{fig:9} to \ref{fig:13}, present, for each of the five periods so defined, the comparative evolution of the number of contaminated recorded by the Ministry and the number of contaminated obtained with the model in $t^m$. 

\begin{figure}[h!]
\hspace{-0.6cm}
 \psfragfig[width=1.15\columnwidth]{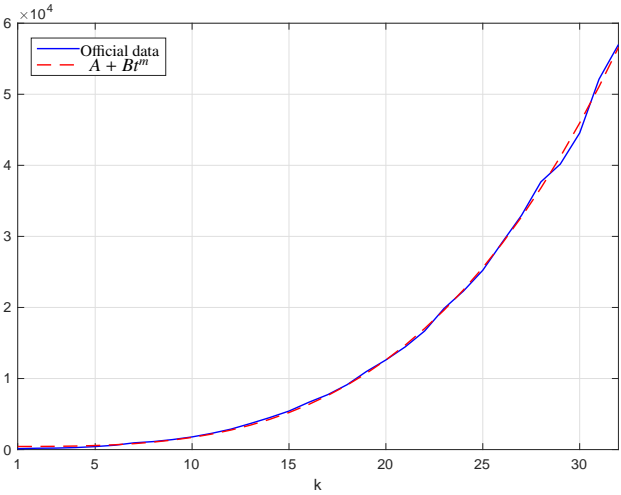}{
  \psfrag{bbbbbbbbbb}[c][][0.7]{\hspace{-0.1cm} $A+Bt^m $}
   \psfrag{aaaaaaaaaa}[c][][0.6]{Official data}
    }
\caption{Comparative evolution, during period 1, of the contaminated recorded  by the Ministry and obtained with the model in $t^m$:whereas $\textrm{mean}(E_r) =-0.0618$ on period 1 for all $k$,  $\textrm{mean}(E_r, k\geq 4) =-0.0059$, which conveys a precision lack of the model at the start (on 3 days) to the first growth phase of the pandemic, as commented in the case of figure \ref{fig:16bis}}
\label{fig:9}
\end{figure}

\begin{figure}[h!]
\hspace{-0.6cm}
 \psfragfig[width=1.15\columnwidth]{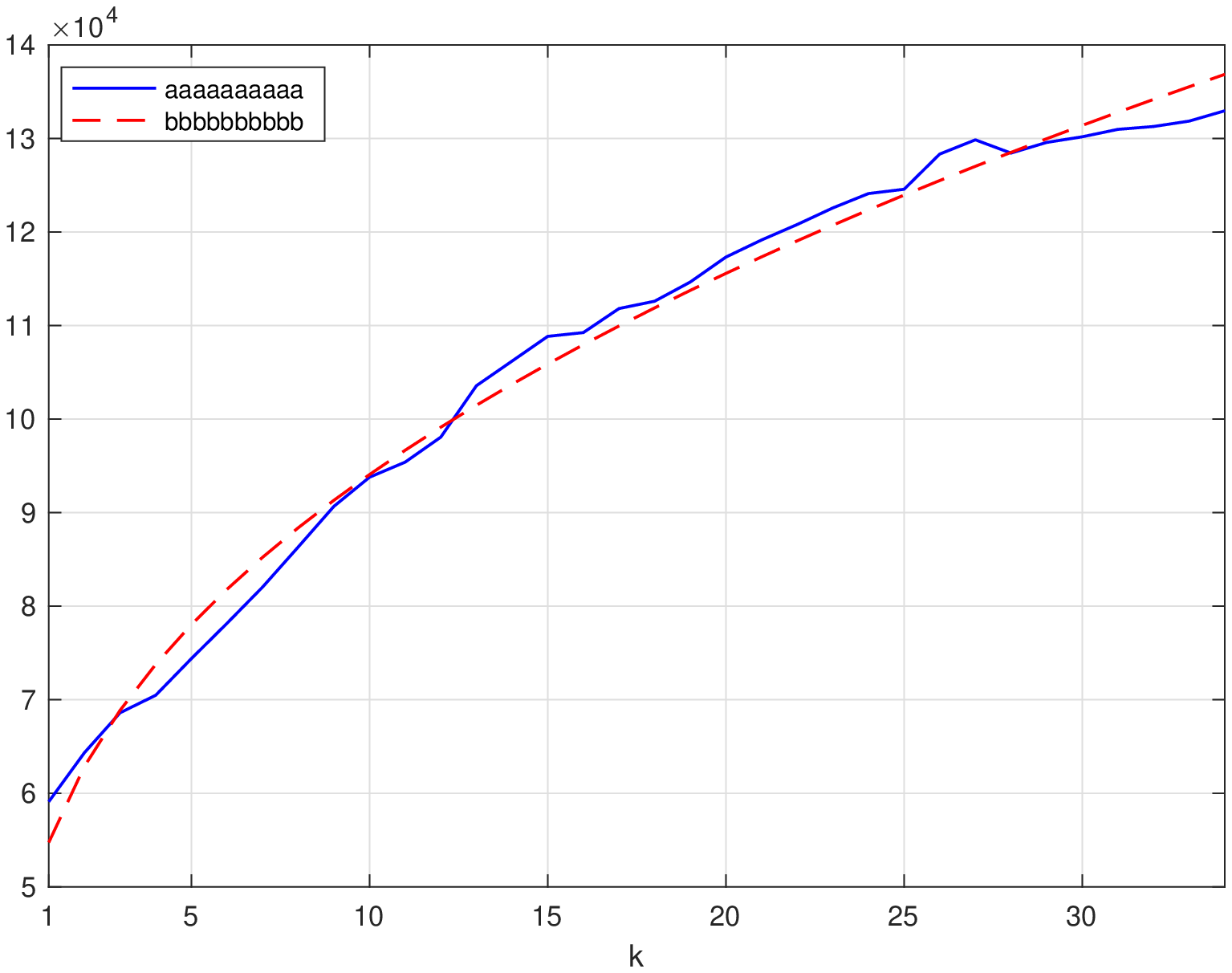}{
  \psfrag{bbbbbbbbbb}[c][][0.7]{\hspace{-0.1cm} $A+Bt^m $}
   \psfrag{aaaaaaaaaa}[c][][0.6]{Official data}
    }
\caption{Comparative evolution, during period 2, of the contaminated recorded  by the Ministry and obtained with the model in $t^m$:  $\textrm{mean}(E_r) =1.5731\, 10^{-4}$} 
\label{fig:10}
\end{figure}

\begin{figure}[h!]
\hspace{-0.6cm}
 \psfragfig[width=1.15\columnwidth]{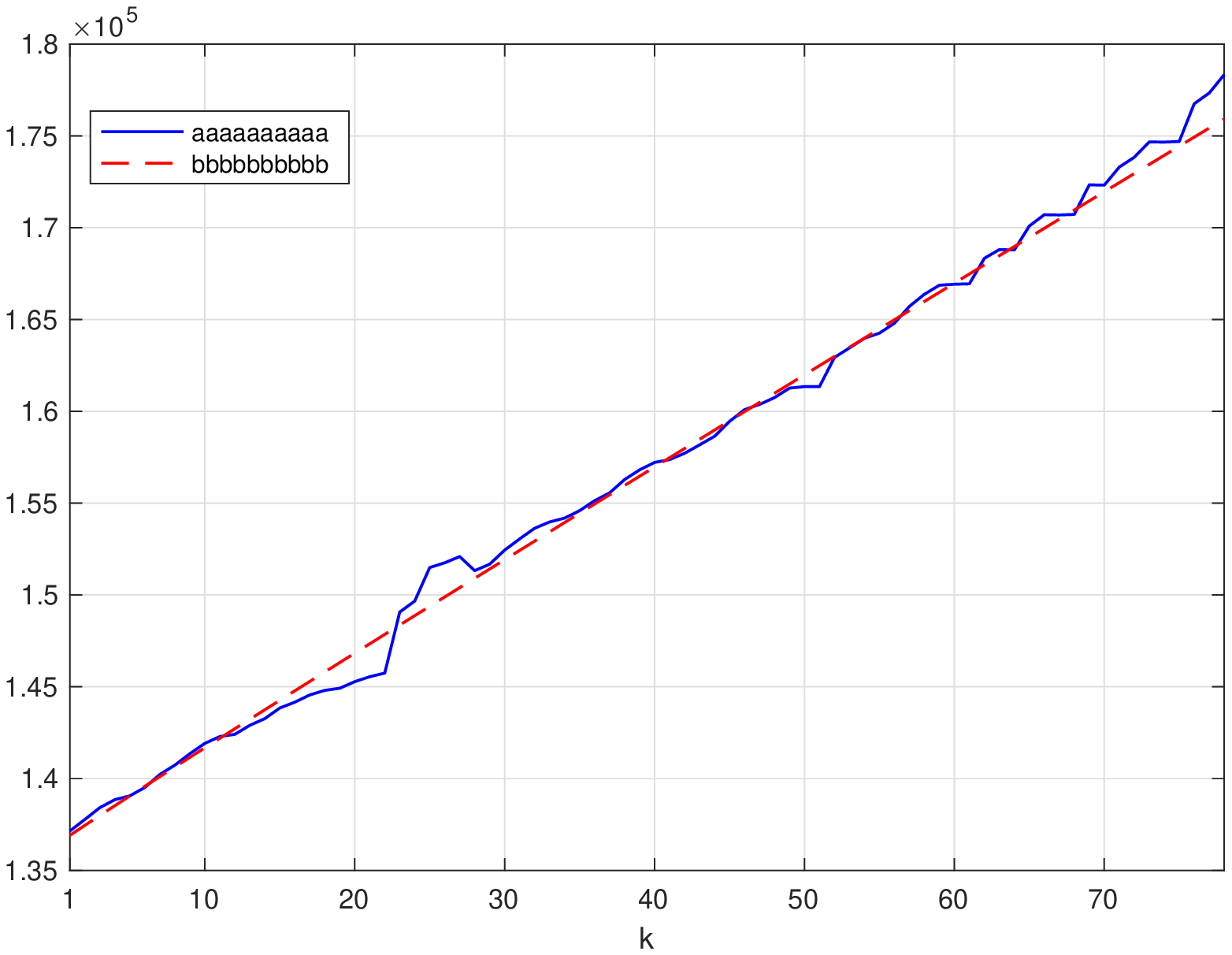}{
  \psfrag{bbbbbbbbbb}[c][][0.7]{\hspace{-0.1cm} $A+Bt^m $}
   \psfrag{aaaaaaaaaa}[c][][0.6]{Official data}
    }
\caption{Comparative evolution, during period 3, of the contaminated recorded  by the Ministry and obtained with the model in $t^m$:  $\textrm{mean}(E_r) =9.4442\, 10^{-4}$}
\label{fig:11}
\end{figure}

\begin{figure}[h!]
\hspace{-0.6cm}
 \psfragfig[width=1.15\columnwidth]{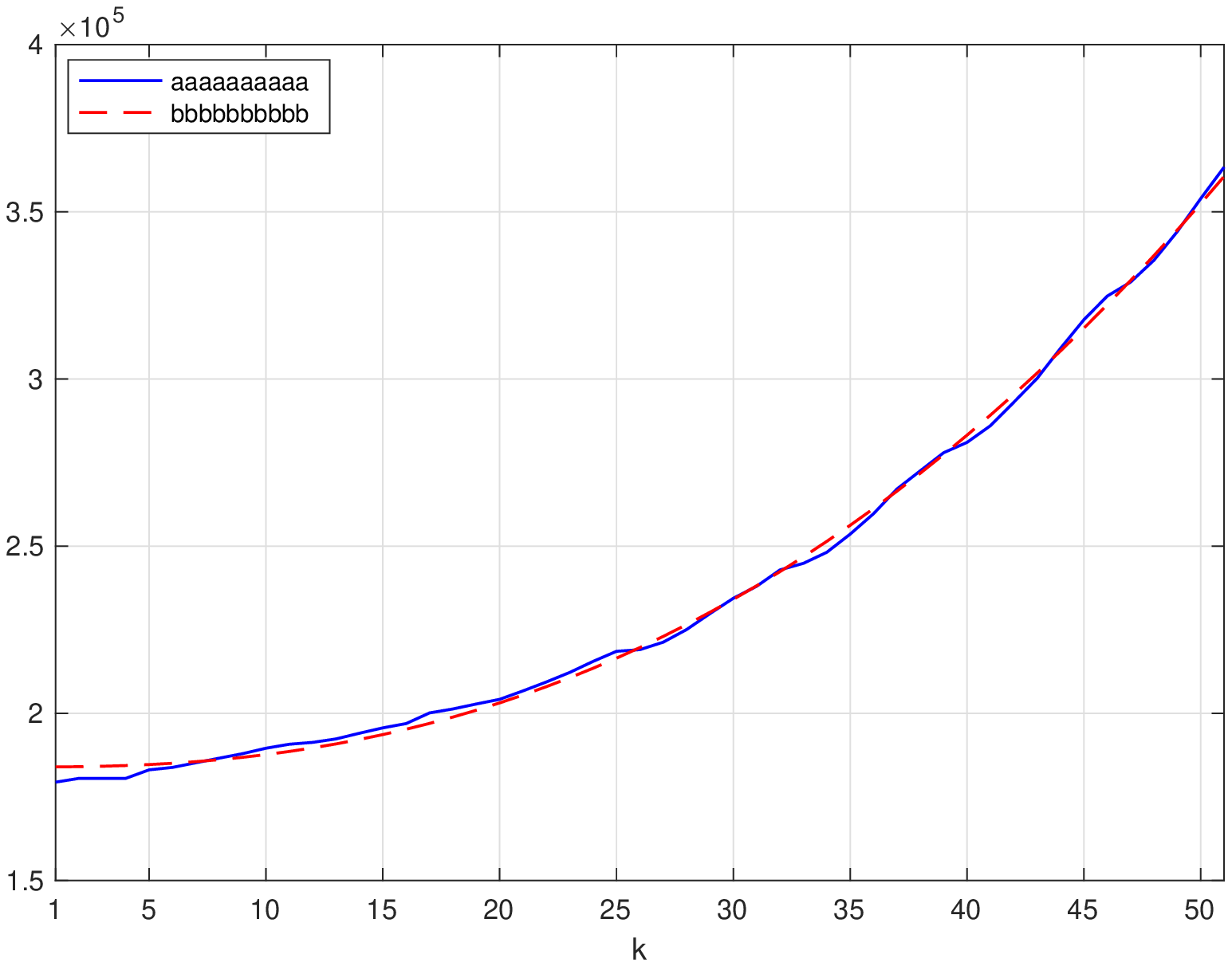}{
  \psfrag{bbbbbbbbbb}[c][][0.7]{\hspace{-0.1cm} $A+Bt^m $}
   \psfrag{aaaaaaaaaa}[c][][0.6]{Official data}
    }
\caption{Comparative evolution, during period 4, of the contaminated recorded  by the Ministry and obtained with the model in $t^m$:  $\textrm{mean}(E_r) =-6.4613\, 10^{-4}$} 
\label{fig:12}
\end{figure}

\begin{figure}[h!]
\hspace{-0.6cm}
 \psfragfig[width=1.15\columnwidth]{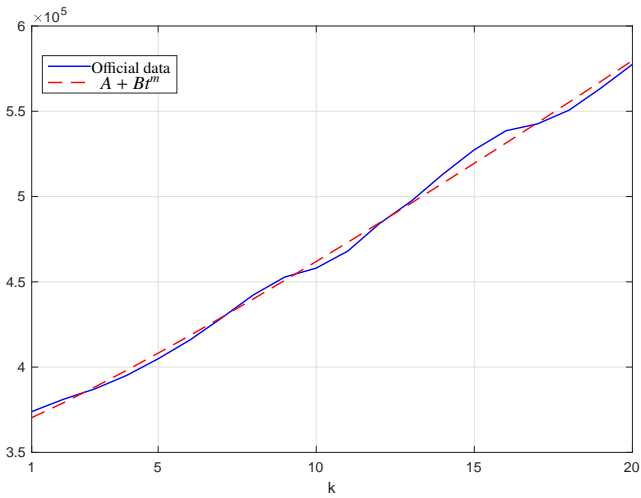}{
  \psfrag{bbbbbbbbbb}[c][][0.7]{\hspace{-0.1cm} $A+Bt^m $}
   \psfrag{aaaaaaaaaa}[c][][0.6]{Official data}
    }
\caption{Comparative evolution, during period 5, of the contaminated recorded  by the Ministry and obtained with the model in $t^m$: $\textrm{mean}(E_r) =7.9843\, 10^{-5}$}
\label{fig:13}
\end{figure}

\section{On the predictivity of the FPM model, after comparing the SIR and FPM approaches}\label{sec:Predictivity_SIR_FPM}

{
In order to better understand how the FPM model is used as a predictor, it is interesting to compare it with the SIR model such as briefly presented in \ref{App:AnnexeSIR}.

The SIR model (with its different variants) is a knowledge and simulation model that uses biological parameters, notably the infection and recovery rates, $\beta $ and $\gamma$. It enables simulating the epidemic evolution through a set of differential equations. Knowing that the parameters $\beta $ and $\gamma$ are not sufficiently known, they can be refined by an identification (or model readjustment) operation. This operation permits to refine the behavior of the SIR model, also by compensating for the model parametric variability problem during the epidemic evolution. 

As for the FPM model, it constitutes a behavior model, whose structure is deduced from a phenomenological analysis (section \ref{subsec:ComplexityInternalStructure}), and whose parameter determination ($A$, $B$ and $m$) requires experimental data describing the epidemic evolution.  
The model originality is due to the parameter $m$ that characterizes the epidemic evolution, by indicating an aggravation or an improvement according to whether it is greater or lesser than one. The three model parameters are obtained thanks to an identification technique that adjusts at best the model behavior in relation to the experimental data. 
It then enables a prediction of the epidemic evolution beyond the identification horizon. If the FPM model is iteratively calculated with data belonging to a sliding window, it can precisely detect any curvature change, and predict the following of this curvature with a precision such as proved in sections \ref{subsec:PredictionLockdownPhase} to \ref{subsec:PredictionVaccinationEllememe}: this advantage is due to the very principle of its determination, according to which it is calculated from the real data and therefore from reality. 
Thus, the SIR model and the FPM model appear, by nature, as complementary models and non competitive. 

In the prediction such as led here with the FPM model, the idea is to use the first effects of a contamination or a sanitary measure (lockdown, vaccination,...) to predict what follows. 
More precisely, the strategy consists in capturing (by the model) the curvature of the first effects, to predict the following as long as the curvature does not change (the conditions then remaining the same).
{{The application of this strategy leads to a prediction technique, whose implementation proves to be particularly simple, because of the simplicity of the FPM model, thus expressing a genuine advantage in favor of this model. }}\label{Rev2:TimeSeries}

\subsection{Predictions justifying a lockdown}\label{subsec:PredictionLockdown}

The model in $t^m$  which is used here as a predictor, is obtained, by identification, from the official data collected before the lockdown, notably between March 1 and March 16 included, March 17 being the first day of the lockdown. The model parameters respectively admit for numerical values:
\[
A = 205.2763,\; B = 1.9419 \;\text{and}\; m= 2.9189.
\]

Using this model until the last day of lockdown, May 10, enables comparing, during the lockdown, between March 17 and May 10 included, the evolution of the prediction and the one of the data provided by the Ministry (figure \ref{fig:14}).

\begin{figure}[t]
\hspace{-0.7cm}
 \psfragfig[width=1.15\columnwidth]{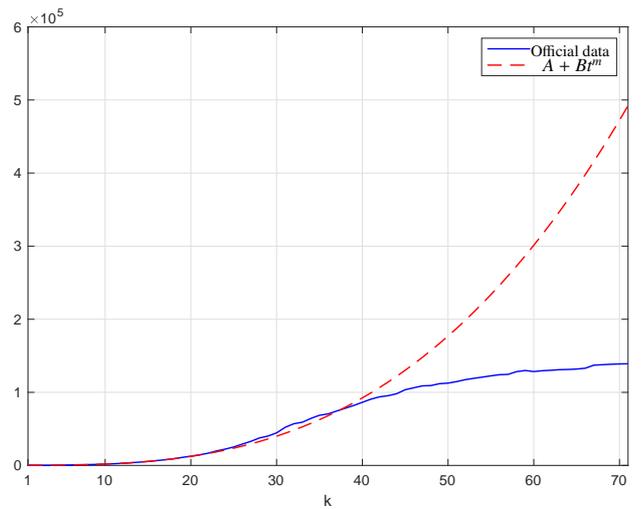}{
  \psfrag{bbbbbbbbbb}[c][][0.7]{\hspace{0cm} $A+Bt^m$}
   \psfrag{aaaaaaaaaa}[c][][0.6]{\hspace{-0.1cm} Official data}
    }
\caption{Comparative evolution, during the lockdown, of the model prediction and the data provided by the Ministry}
\label{fig:14}
\end{figure}

The comparison which is assessed on May 10, namely $492\,$ $274$ cases predicted without the lockdown and $139\, 063$ cases confirmed with the lockdown, turns out to be very eloquent in the sense that it well expresses the interest of the lockdown. 

Concerning the precision associated with such a prediction, it is not possible to verify it as the real data stemming from the Ministry are relative to the lockdown. Moreover,  it is difficult to provide an estimation of this precision, as the prediction model is the subject of an exploitation that is  demanding to say the least: indeed, it is established on a period limited  to 16 days, whereas it is used with a prediction horizon of 55 days; the opposite would have been more in favor of the precision. 

\subsection{Verified predictions in lockdown phase}\label{subsec:PredictionLockdownPhase}
The idea is to use the significative data of the first lockdown effects, to calculate the parameters of the model which is then used as a predictor, to predict the contaminated total number at the end of the lockdown. As the first effects appear only after ten days, in order to capture at best these first effects, the most appropriated data interval has been considered. This interval starts after the twelfth day of lockdown to ensure a supplementary 2 day margin beyond the ten first days; this margin aims at guaranteeing significative values to the interval first data. Moreover, in order to benefit from a sufficient number of significative data, a nine day interval has been chosen. This interval thus corresponds to 2 supplementary days in relation to a one week interval, which has been the most used to calculate the model. 

So, for the lockdown between March 17 to May 10 2020, the data interval used to compute the model is from March 29 to April 6 (see table \ref{tb:Ident2020}), and the prediction horizon is from April 7 to May 10, therefore of 34 days, namely five weeks up to a day. 
With the data of 9 days, from March 29 to April 6, the obtained model is such that: 
\[A=27204.97, \; B=12404.92 \; \text{and} \; m=0.6082431.\] 
Its representativity is confirmed by a low modeling error, as the relative error mean is only of $0.0009 \%$.
For the prediction horizon of 34 days, from April 7 to May 10 (see table \ref{tb:Prediction2020}), the model predictivity is confirmed by an absolutely admissible prediction error  for a political decision-maker. It is indeed less than $10 \%$, as the modulus maximum of the relative errors, obtained at the last day, equals $6.93 \%$ (the mean of these errors being of $0.71 \%$): this value therefore gives the prediction error at May 10, which well numbers the precision of the sought prediction at this day.

\subsection{Verified predictions in vaccination phase}\label{subsec:PredictionVaccinationPhase}
To prove that the model validity is kept while an increase of vaccination covering, the model in vaccination phase has been tested, and this, not only as for its representativity, but also as for its predictivity. 

The model is calculated with the data collected on one week (see table \ref{tb:Ident2021}), notably the one from April 26 to May 2 2021, which is located in a high increase of vaccination covering. The model is then used as a predictor on a prediction horizon of six weeks, namely 42 days, from May 3 to June 13. 
With the data of the week from April 26 to May 2 2021, the obtained model is such that:
\[A=5436574.3, \;B=65600.93\; \text{and} \; m=0.6229530.\]
Its representativity is proved by a low modeling error, as the modulus maximum of the relative errors is only of $ 4.82 \, 10^{-6}\%$.  
For the prediction horizon of 42 days, from May 3 to June 13 (see table \ref{tb:Prediction2021}), the model predictivity is proved by a low prediction error, as the modulus maximum of the relative errors, obtained at the last day, equals $1.18 \%$, and the mean of these errors is of $-0.23 \%$. Note that at the end of the fifth week, the maximum and mean only equal $0.54 \%$ and $0.09\%$.

\subsection{Verified predictions on the vaccination itself}\label{subsec:PredictionVaccinationEllememe}
This section concerns the vaccination as a complex phenomenon, and notably the representation and prediction of its evolution by a non-integer model. It well illustrates and makes concrete section \ref{subsec:GlobalBehaviorProPhenContamination}, which stresses that a model in $t^m$ is a good modeling tool to represent complexity, beyond the one of a viral spreading. In this new context, the considered data turn on the accumulated number of the totally vaccinated individuals.

To study the representativity of the model in $t^m$, the data collected on 90 days, from March 29 to June 26 2021, correspond to a high increase of vaccination. By decomposing  this period of 90 days in two identical periods of 45 days, it is possible to calculate three models: the first one for the period from March 29 to May 12, the second one for the period from May 13 to June 26, and the third one for the whole study period from March 29 to June 26. Thus, the first model is such that: 
\[A=2900427.3, \; B=19631.06\; \text{and} \; m=1.511465.\] 
The second one is such that:
\[A=9522593.4,\; B=22512.59 \; \text{and}\;  m=1.635698.\]
The third one is such that: 
\[A=3238558.7,\; B=10477.08\;   \text{and}\;  m=1.640339.\]
The representativity of each of these models is indicated by a low modeling error, as shown by figures \ref{fig:Pred2903_1205}, \ref{fig:Pred1305_2606} and \ref{fig:Pred2903_2606}: more precisely, the relative error mean equals $0.072\%$ and $0.0026 \%$ for the first and second models, and $-0.14 \%$ for the third one, a value which is of course greater given the period duration, but which is not explosive.
To study the predictivity of the model in $t^m$, the data are collected on one week (see table \ref{tb:IdentVaccines2021}), from May 8 to May 14, and also correspond to a high increase of vaccination, this week being located around the middle of the period from March 29 to June 26. With the data of this week, the obtained model is such that:
\[A=8051645.3,\; B=206316.7\; \text{and} \; m=0.9995548.\]
Its representativity is expressed by a low modeling error, as the relative error mean is only of $  -1.4022\,10^{-4} \%$.
For a prediction horizon of four weeks, namely 28 days, from May 15 to June 11 (see table \ref{tb:PredictionVaccines2021}), the model predictivity is expressed by a low prediction error: even if the modulus maximum of the relative errors reaches $6.86 \%$ at the third day of the second week of the prediction horizon, the mean of these errors is only of $-3.57 \%$.

\begin{figure}[t]
\hspace{-0.7cm}
 \psfragfig[width=1.15\columnwidth]{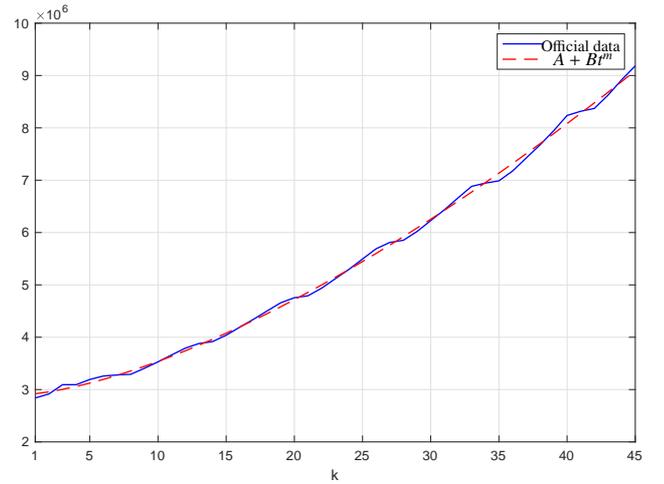}{
  \psfrag{bbbbbbbbbb}[c][][0.7]{\hspace{0cm} $A+Bt^m$}
   \psfrag{aaaaaaaaaa}[c][][0.6]{\hspace{-0.1cm} Official data}
    }
\caption{Comparative illustration, for vaccination, of the official data and the predictions from March 29 to May 12 (45 days): $\textrm{mean}(E_r)=7.1706\,10^{-4}$}
\label{fig:Pred2903_1205}
\end{figure}

\begin{figure}[t]
\hspace{-0.7cm}
 \psfragfig[width=1.15\columnwidth]{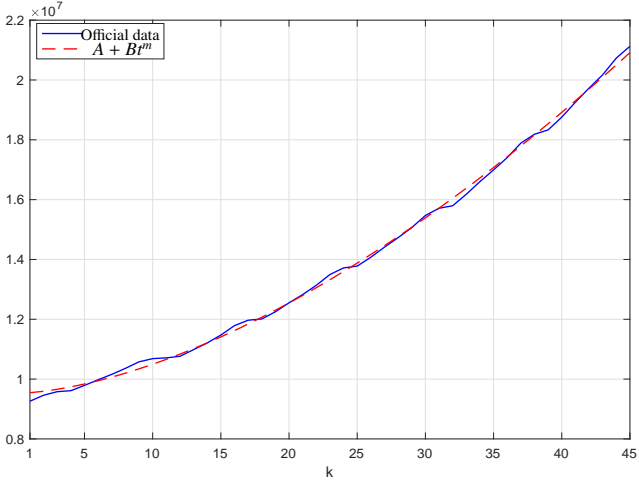}{
  \psfrag{bbbbbbbbbb}[c][][0.7]{\hspace{0cm} $A+Bt^m$}
   \psfrag{aaaaaaaaaa}[c][][0.6]{\hspace{-0.1cm} Official data}
    }
\caption{Comparative illustration, for vaccination, of the official data and the predictions from May 13 to June 26 (45 days): $\textrm{mean}(E_r)=2.6055\,10^{-5}$}
\label{fig:Pred1305_2606}
\end{figure}

\begin{figure}[t]
\hspace{-0.7cm}
 \psfragfig[width=1.15\columnwidth]{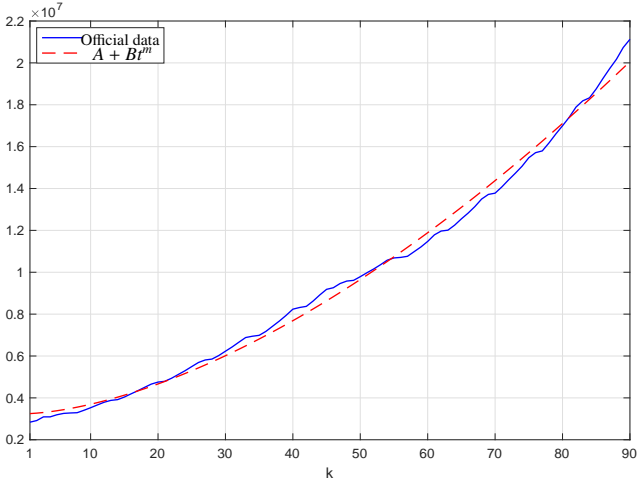}{
  \psfrag{bbbbbbbbbb}[c][][0.7]{\hspace{0cm} $A+Bt^m$}
   \psfrag{aaaaaaaaaa}[c][][0.6]{\hspace{-0.1cm} Official data}
    }
\caption{Comparative illustration, for vaccination, of the official data and the predictions from March 29 to June 26 (90 days): $\textrm{mean}(E_r)=-1.4462 \, 10^{-3}$}
\label{fig:Pred2903_2606}
\end{figure}

}

\section{Conclusion}\label{sec:conclusion}

{{This article, which aims at well capturing the time series and particularly those of the contaminations, develops an original approach of a very simple deterministic model with three parameters, the FPM model,}} whose non-integer power is significative to the curvature of the viral spreading evolution. This model enables representing at each day the totality of the contaminated, and the internal structure, founded on a geometric sequence ``with variable ratio'', enables, on one hand, to formalize the daily censing of the new contaminated and, on the other hand, to represent day after day the whole of the new contaminated. 


The simplicity of the model is due to its variation in $t^m$ which is a response of the non-integer differentiation or integration operator, an operator known for its capacity to simply represent complex phenomena such as diffusion phenomena (of thermal origin (\cite{Battaglia2000}) or fluidic one (\cite{Oustaloup2014})).

Founded on a known structure and unknown parameters, $q_k$, to be determined, the construction point by point of a non-integer time function of the same form as the model in $t^m$, can be interpreted as a \emph{time synthesis} in non-integer. This  synthesis is not without evoking an analogy with the \emph{frequency synthesis} of the non-integer differentiation or integration operator, from a synthesis network having a known structure and unknown parameters to be determined. Whilst there is indeed an analogy through a common context of synthesis, this analogy ends here, in the sense that both syntheses in question are not of the same nature. On one hand, the first synthesis turns on a time response of a non-integer operator, while the second one turns on the frequency response of the operator itself. On the other hand, the first synthesis uses an evolutive structure of synthesis, due to an \emph{evolutive dimension} that increases with time, whereas the second one uses a fixed structure of synthesis with a fixed given dimension. 

The model internal structure is conceived so as to introduce ratios likely to represent those of the new contaminated between two consecutive days, then used without seeking to represent the unpredictable inherent to reality hazards, particularly the random fluctuations of these ratios. It is true that, as shown in the article, these fluctuations have a negligible effect on the slower dynamics to be modelled by the model in $t^m$, and this, because of the self-filtering character of the internal structure, highlighted through a filtering double effect. More generally, this self-filtering character removes the paradox between the simplicity of a model in $t^m$ and the reality complexity, thus consolidating the interest of this type of model in modeling complex systems and phenomena. 


Used as a prediction model, to predict the number of cases liable to be affected by the virus, the FPM model highly justifies the lockdown choice, without which the number of contaminated would have really exploded.
{{After comparing the FPM model with the SIR model (section \ref{sec:Predictivity_SIR_FPM} and \ref{annex:B}), its predictivity is studied through its prediction precision, and is confirmed by the verified predictions in lockdown phase, in vaccination phase, and even for the vaccination itself; its simplicity enables a very simple implementation of the prediction technique.}} \label{rev1:SIRQ2}

Even if the results are briefly presented in annex for the article length sake, the model in $t^m$ has also been applied with success to the hospitalized, with or without intensive care, and even to the deceased (\ref{App:Annexe2}). 
For all the hospitalized, the model can present negative parameters, knowing that the model is then supposed to represent the daily data, and no longer the data sum. 

In order to better measure the contribution of the article, in \ref{sec:modelcomparison}, the proposed model in $t^m$ is successively compared to two known models having the same number of parameters, namely a ``convexity model'' and a ``convexity and concavity model'', the  Verhulst model, which has been rewritten under the form of a model in $m^t$ so as to introduce a duality with the proposed model. The comparison is eloquent, in the sense that a quick analysis of the comparative illustration of the representativity of the models, clearly proves to be in favor of  the model in $t^m$, whose representativity is better or more general according to the compared model. 

Furthermore, in a context that scientifically involves complexity and simplicity, the self-filtering action of the internal structure, makes it possible to show the compatibility between the \emph{internal complexity} (that the internal structure and its stochastic behavior present) and the \emph{global simplicity} (that the determi\-nistic model offers via its power-law $t^m$): it is true that the non-integer character of the power very simply expresses or represents complexity.



\appendix

\section{Some applicative fields in which power-law dynamics occurs}\label{App:AnnexePower-law}
Power-law dynamics such as expressed by $t^m$ can be found in many applicative fields.

For example, in cosmology, the expansion of the universe is explained in \cite{Frieman2008} through the cosmic scale factor, $a(t)$, which is controlled  by the dominant energy form $a(t) \propto  t^{\frac{2}{3(1+w)}}$ for a constant $w$.

Also in fluid dynamics, the Tanner law, which was established in \cite{Tanner1979} for explaining the spreading of silicone oil drops, enables explaining the spreading dynamics of fluid droplets according to $R(t) \propto t^m$ where $R(t)$ is the radius of the droplet (see \cite{Liang2009}).

And even in neurobiology, the dynamics of biological systems appear to be exponential over short-term courses and to be in some cases better described over the long-term by power-law dynamics. A model of rate adaptation at the synapse, between inner hair cells and auditory-nerve fibers, is presented in \cite{Zilany2009}. 



A non-exhaustive list of application fields that use the po\-wer-law dynamics, can be cited: in biochemical systems (\cite{Savageau1970}), in DNA dynamics (\cite{Andreatta2005}), in semiconductor nanocrystals (\cite{Sher2008}), in water ressour\-ces (\cite{Harman2009}), in oscillators (\cite{Korabel2007}) and even  in financial market (\cite{Gabaix2003}), etc.  
\vspace{-0.5cm}


{
\section{On the SIR model}\label{App:AnnexeSIR}\label{annex:B}
In viral epidemiology, the total population is divided into several disjoint sets, called compartments. The population of a compartment evolves  according to time and the links between the compartments. An individual of a compartment at an instant $t$ stems from another compartment among other ones, and can then become member of another compartment among other ones. In the SIR method, there are, at each instant, three compartments: \emph{Susceptible} (S) gathers the non-contaminated individuals that are non immune and susceptible to catch the virus. \emph{Infected} (I) individuals who have been infected and are capable of infecting susceptible individuals. Finally, individuals, who have been infected enter the \emph{Removed} (R) compartment, through either recovery or death.

An individual from S at an instant $t$ can become member of I at an instant $t'>t$, then member of R at an instant $t''>t'$. Let $\beta$ be the infection rate (or virus transmission rate), which corresponds to the rate of susceptible individuals at instant $t$ that become infected at instant $t+\Delta t$. Let $\gamma$ be the recovery rate, which corresponds to the rate of infected individuals at instant $t$ that recover at instant $t+\Delta t$. It is assumed that these rates stay constant during the considered period. Let $S(t)$, $I(t)$ or $R(t)$ be the total number of individuals in the corresponding compartment at instant $t$. Also assume that the total number $N$ of the population stays constant during the considered time period, therefore $S(t)+I(t)+R(t)=N$. Also assume that $\Delta t$ remains constant. In practice, it corresponds to a day as the data are collected daily.  

The following equations are obtained:
\begin{equation}
\left\{
\begin{array}{ccc}
S(t+\Delta t) - S(t) &= &-\displaystyle \frac{\beta}{N}  \Delta t S(t) I(t)\\
I(t+\Delta t) - I(t) &=  &\displaystyle\frac{\beta}{N} \Delta t S(t) I(t) - \gamma \Delta t I(t)\\
R(t+\Delta t) - R(t) &= &\gamma\Delta t I(t),
\end{array}\right.
\end{equation}
these equations resulting from the discretization of the three differential equations:
\begin{equation}
\left\{
\begin{array}{ccc}
\frac{d S(t)}{dt} &= &- \displaystyle \frac{\beta}{N}  S(t) I(t) \\
\frac{d I(t)}{dt} &=  &\displaystyle \frac{\beta}{N}   S(t) I(t) - \gamma I(t)\\
\frac{d R(t)}{dt} &= &\gamma I(t).
\end{array}\right.
\end{equation}

\emph{Remark 1} - The ratio $R_0 = \beta/\gamma$ is the basic reproduction number: if $R_0>1$, the epidemic progresses; if $R_0<1$, the epidemic regresses.

\emph{Remark 2} - The presented model is the basic SIR model which is the subject of numerous improvements, through seve\-ral derived compartment models, such as SIS, SIRD, SIRV, MSIR, SEIS, SEIR, MSEIR, MSEIRS, etc.
}

\section{Application of the FPM model to the official data on the hospitalized and the deceased}\label{App:Annexe2}
\begin{figure}[h]
\hspace{-0.6cm}
 \psfragfig[width=1.15\columnwidth]{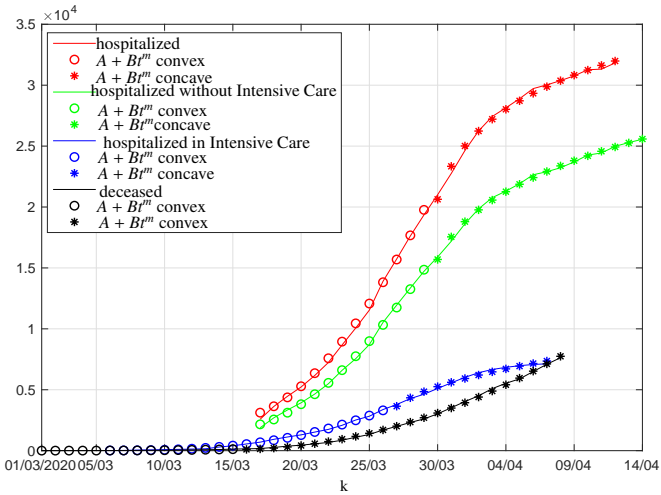}{
  \psfrag{k}[c][][0.6]{\hspace{0cm} k}
  \psfrag{aaaaaaaaaa}[c][][0.65]{\hspace{0cm} hospitalized}
   \psfrag{bbbbbbbbbb}[c][][0.65]{\hspace{0.6cm}$A+B t^m$ convex}
  \psfrag{cccccccccc}[c][][0.65]{\hspace{0.8cm} $A+B t^m$ concave}
   \psfrag{ddddddddddddddddddddddddddddddd}[c][][0.65]{\hspace{0.05cm} hospitalized without Intensive Care} 
    \psfrag{eeeeeeeeee}[c][][0.65]{\hspace{0.6cm} $A+B t^m$ convex}
   \psfrag{ffffffffff}[c][][0.65]{\hspace{1.4cm} $A+B t^m$concave}
     \psfrag{gggggggggg}[c][][0.65]{\hspace{2.8cm} hospitalized in Intensive Care}
   \psfrag{hhhhhhhhhh}[c][][0.65]{\hspace{0.6cm} $A+B t^m$ convex}
     \psfrag{iiiiiiiiii}[c][][0.65]{\hspace{1.6cm} $A+B t^m$ concave}
     \psfrag{jjjjjjjjjj}[c][][0.65]{\hspace{0.6cm} deceased}
   \psfrag{kkkkkkkkk}[c][][0.65]{\hspace{0.8cm} $A+B t^m$ convex}
     \psfrag{llllllllll}[c][][0.65]{\hspace{1.5cm} $A+B t^m$ convex}
    }
\caption{Comparative evolution, during periods 1 and 2, of the hospitalized (\textcolor{red}{$––$}), hospitalized without intensive care (\textcolor{green}{$––$}),  hospitalized in intensive care (\textcolor{blue}{$––$}), and deceased (\textcolor{black}{$––$}) recorded by the Ministry and obtained with the models in $t^m$, one for the convex part (o) and one for the concave part ($*$)}
\label{fig:17}
\end{figure}

The proposed model in $t^m$, $A+Bt^m$, has also been applied with success to the hospitalized, with or without intensive care, and even to the deceased cases for the first two periods, that is to say the first convex period and the first concave one. The comparative evolutions are illustrated on figure \ref{fig:17}, where the official data are plot in solid lines, the models in the convex parts in circle symbols and the models in the concave parts are in star symbols. As expected, the provided models well fits the official data. Moreover, the parameters obtained after identification for the different models for each period  are provided in table \ref{tb:annexe}, and the maximum relative error is provided, thus validating the proposed models. 
\vspace{0cm}

\begin{table*}[h]
\begin{center}
\captionof{table}{Identification parameters for the models in $t^m$ of the hospitalized, hospitalized without intensive care, hospitalized in intensive care, and deceased for period 1 (convex part) and for period 2 (concave part)} \label{tb:annexe}
\begin{tabular*}{\textwidth}{@{\extracolsep{\fill}} c|c|c|c|c|c|c|c}
\toprule
Data    & Dates 				& Part	 &  $k$	& $m$ & $A$ & $B$ & $\textrm{mean}\left ( E_r \right)$\\
\hline \hline
\multirow{2}{*}{hospitalized}& $17/03-29/03$ &convex part	& $1-13$ & $1.6523$		& $2876.401$		&  $243.8106$	& $-4.4\, 10^{-4}$\\
			   & $30/03-12/04$ &concave part	& $1-14$ & $0.0964$		& $-18558.02$		&  $39190.75$	& $1.4\, 10^{-5}$\\
\hline	
hospitalized without & $17/03-29/03$ &convex part	& $1-13$ & $1.6402$		& $1956.690$		&  $ 191.9422$	& $-6.4\, 10^{-4}$\\
	intensive care		   & $30/03-15/04$ &concave part	& $1-16$ & $0.2572$		& $6290.149$		&  $9504.363$	& $4.7\, 10^{-6}$\\
\hline	
hospitalized in& $06/03-26/03$ &convex part	& $1-21$ & $2.8455$		& $14.7447$		&  $0.5692$	& $1.5\, 10^{-2}$\\
intensive care   & $27/03-07/04$ &concave part	& $1-12$ & $0.4078$		& $ 1534.7680$		&  $2113.590$	& $1.0\, 10^{-4}$\\
\hline	
\multirow{2}{*}{deceased}& $01/03-15/03$ &convex part	& $1-15$ &  $3.2097$		& $3.1063$		&  $0.0200$	& $-3.9\, 10^{-2}$\\
			   & $16/03-09/04$ &convex part	& $1-27$ & $1.9951$		& $88.4450$		&  $13.5012$	& $3.8\, 10^{-2}$\\
\bottomrule	
\end{tabular*}
\end{center}
\end{table*}

{{
\section{Comparison of the representativity of the FPM model and of two known models with three parameters} \label{sec:modelcomparison}
For a model, representativity (measured by the degree of representativity), is a characteristic that conveys the model capacity to represent reality. In this section, reality is materialized by the evolution of all the contaminated counted each day during period 1; the representativity of this reality is illustrated, in a comparative way, for the model in $t^m$ and for two known models having also three parameters. If the model in $t^m$ can be interpreted as a ``convexity or concavity model'', the two known models considered here can be interpreted as a ``convexity model'' for the one and as a ``convexity and concavity model'' for the other, a model that we have rewritten under a form in $m^t$, dual to the one of the model in $t^m$.

\subsection{The model of Verhulst: a convexity and concavity model (or sigmoid model)}
Taking into account  a convex then concave growth, the Verhulst model and its generalizations are liable to represent,  both first and second growth phases of the epidemics.

Around 1838, P.F. Verhulst (\cite{Verhulst1838,Verhulst1845}) has established a model of dynamical behavior of a popu\-lation, named logistic growth model. If $a_k$ is the size of the popu\-lation at day $k$, for $k\geq 0$, the growth rate of this population at day $k+1$ is $(a_{k+1}-a_k)/a_k$. 

This rate is equal to the difference between the birth rate and the death rate at day $k$. 
Assuming that the daily birth rate and daily death rate are affine functions of the population size, it then comes:
\[
\frac{a_{k+1}-a_k}{a_k} = N(a_k)-M(a_k),
\]
with:
\[
N(x)=n_1+n_2 x \quad \text{and}\quad M(x) = m_1 + m_2 x.
\] 

This expression is none other than the discretization of a time differential equation in $f(t)$, where $f(k)=a_k$, namely:
\[
f'(t) = f(t)\left[ N\left(f(t)\right) - M\left(f(t)\right)\right] \quad \text{with}\quad f(0) = a_0,
\]
therefore, finally:
\[
f'(t) =a f(t)\left [1-  \frac{f(t)}{K} \right],
\]
by writting:
\[
a = n_1- m_1 \quad \text{and} \quad  K = \frac{n_1- m_1}{m_2 -n_2}= \frac{a}{m_2 -n_2}.
\]

The solution of this differential equation, $f(t)$, admits an expression with three parameters, $K$, $a_0$ and $a$:
\begin{equation}\label{eq:Verhulst}
f(t) = \frac{K}{1+\left(\frac{K}{a_0}-1 \right)e^{-at}},
\end{equation}
an expression which is easier to verify than to establish. 
%

\subsection{Use and generalization of Verhulst model}
Many articles use this model with three parameters for the COVID-19. 
\cite{Zhou2020}, \cite{Zhao2020}, \cite{Remond2020}, use the same differential equation.
\cite{Kyurkchiev2020} use a model which is generalized by:
\[
y(t) = \frac{r}{1+e^{a_0 + a_1 t+a_2 t^2+\ldots }}.
\]

%

There are several possible generalizations of Verhulst model such as proposed by \cite{Roosa2020} who give two well-known generalizations with four parameters:
\begin{itemize}
\item the generalized logistic model (GLM), namely
\[
\frac{dC(t)}{dt} = rC(t) ^p \left( 1-\frac{C(t)}{K}\right),
\]
used, among others, by \cite{Viboud2016} and \cite{Ganyani2018}, the four parameters being $r$, $p$, $K$ and $C(0)$;
\item the Richards model, namely
\[
\frac{dC(t)}{dt}  = rC(t) \left( 1-\left (\frac{C(t)}{K}\right)^a \right),
\]
used, among others, by \cite{Wang2012} and \cite{Richards1959} in 1959, the four parameters being $r$, $a$, $K$ and $C(0)$.
 \end{itemize}
 
Richards model is found under a generalized form with five parameters, called generalized Richards model (GRM), namely (\cite{Wu2020}):
\[
\frac{dC(t)}{dt} = rC(t) ^p \left( 1-\frac{C(t)}{K}^a\right),
\] 
the five parameters being $r$, $p$, $a$, $K$ and $C(0)$, with $p\in  [0,1]$.

\subsection{A convexity model}
The model presented  in this section is used to represent the first growth phase of the epidemics. 
Its use is indeed kept for convex growths. 

By only taking the first factor of the  GLM and GRM models, these models are reduced to a same model such as  presented in \cite{Wu2020}, \cite{Viboud2016} and  \cite{Tolle2003}, namely:
\begin{equation}\label{eq:ConvexModel}
\frac{dC(t)}{dt}  = rC(t)^p \quad  \text{with}\quad r>0 \,\text{and} \, p\in[0,\,1],
\end{equation}
a differential equation that defines a model with three parameters, $r$, $p$ and $C(0)$, and that admits three different solutions according to whether $p=0$, $p=1$ or $0<p<1$: an affine solution for $p=0$, an exponential solution for  $p=1$ or a sub-exponential solution for $0<p<1$.

If for $p=0$, the solution is of constant slope, $r$, as $C(t)  = rt+C(0)$, for $p=1$, the differential equation then being linear, its solution is here the one of a system of first order, namely:
\[
C(t) = C(0) e^{rt}.
\]

For  $0<p<1$, the differential equation then being  nonlinear, the establishment of its solution requires a rewriting of the  equation in conformity with:
\[
C(t)^{-p}  C'(t)  = r,
\]
namely:
\[
\left(C(t)^{1-p}\right)' = r(1-p),
\]
or, by integrating, thus introducing  an integration constant  $C$:
\[
C(t)  ^{1-p} = r(1-p) t+C,
\]
therefore, finally:
\[
C(t) = \displaystyle\left( r(1-p)t +C(0) ^{1-p} \right)^{\frac{1}{1-p}},
 \]
a solution  with  non-integer exponent, that we rewrite here under a form comparable to the one of the model in $t^m$, $A+Bt^m$, namely:
\[
C(t) = \left ( A'+B't  \right)^{m'},
\]
with:
\[
A' = C(0)^{1-p}, \quad  B' = r(1-p)
\]
and
\[
m' = \frac{1}{1-p}> 1 \quad \text{as} \quad 0<p<1.
\]

Given the two model structures, if the model in $t^m$ can be interpreted as a ``model with explicit power of time'', the other model can be interpreted as a ``model with implicit power of time'', a  power which is  non-integer for both models.

%

Even if the model represented by the differential equation \eqref{eq:ConvexModel} imposes, on its solution $C(t)$, an exponent $m'$ greater than 1, we have wished to find out more, by studying this solution for an exponent $m'$ comprised between 0 and 1, which corresponds to $p<0$ for the differential equation and its solution. 

Thus, to lead a comparative study between the parameters of both models for $m$ and $m'$ between 0 and 1, and this for simply understanding why the ``convexity model'' is not  proposed for concave growths, it is convenient to approximately express $A'$ and $B'$ versus $A$, $B$, $m$ and $m'$ by considering the identity of the models at an initial instant, $t=0$, and at a final instant, $t=t_n$, then to analyze $A'$ and $B'$ for the small values of $m'$.

Let be the two models, denoted here by $C_1(t)$ and $C_2(t)$:
\[
C_1(t) = A+Bt^m,\quad \text{with}\quad  A,\,  B>0 \quad \text{and} \quad 0<m<1,
\]
and 
\[
C_2(t) = \left(A'+B't\right)^{m'},\quad \text{with}\quad  A',\,  B'>0 \quad \text{and} \quad 0<m'<1.
\]

For $t=0$:
\[
C_1(0) = A \quad \text{and}\quad C_2(0) = A'^{m'},
\]
then, given  that  $C_1(0) = C_2(0)$:
\begin{equation}\label{eq:Aprime}
A' =  A^{\frac{1}{m'}}.
\end{equation}
 
For $t=t_n$:
\[
C_1(t_n) = A+Bt_n^m \quad \text{and}\quad C_2(t_n) = (A'+B't_n)^{m'},
\]
then, given  that  $C_1(t_n) = C_2(t_n)$:
\[
A' +B't_n= (A+Bt_n^m)^{\frac{1}{m'}},
\]
from where one draws:
\[
B' = t_n^{-1} \left[ (A+Bt_n^m)^{\frac{1}{m'}} - A^{\frac{1}{m'}}   \right],
\]
namely, by writting 
\[
Btn^m = \alpha A\quad \text{with}\quad \alpha>0:
\]
\begin{equation}\label{eq:Bprime}
B' = t_n^{-1} A^{\frac{1}{m'}} \left[ (1+\alpha)^\frac{1}{m'} -1 \right]. 
\end{equation}

Given that $A$, $\alpha$ and $m'$ are positive, the relations  \eqref{eq:Aprime} and \eqref{eq:Bprime} well show that  $A'$ and  $B'$ tend towards infinity  when $m'$ tends towards zero. This phenomenon is in conformity with the huge values of  $A'$ and  $B'$ that we have obtained, even for $m'=0.308$ which is far from zero, in the case of an identification attempt of the second period with $m'$  between 0 and 1. Indeed, whereas $A=30809.42$ and $B=23934.54$ ($m$ being $0.422$), the values  of $A'$ and $B'$ reach respectively, for a similar precision, $A' =  0.106\, 10^{16}$ and $B'   = 0.123 \, 10^{16}$  (for $m' = 0.308$).

\begin{figure}[h]
\hspace{-0.6cm}
 \psfragfig*[width=1.15\columnwidth]{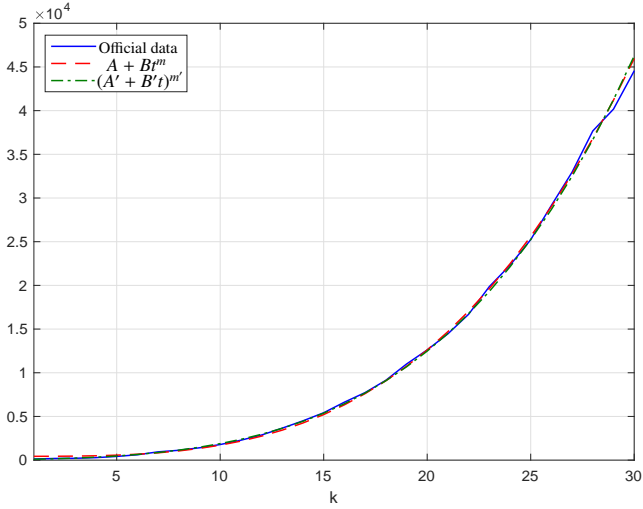}{
  \psfrag{bbbbbbbbbb}[c][][0.7]{\hspace{0cm} $(A'+B't)^{m'}$}
    \psfrag{cccccccccc}[c][][0.7]{\hspace{0cm} $A+Bt^{m}$}
   \psfrag{aaaaaaaaaa}[c][][0.6]{\hspace{-0.1cm} Official data}
    }
\caption{Comparative evolution, during period 1, of the contaminated recorded by the Ministry, and obtained
with the model in $t^m$, $C_1(t)$, and the ``convexity model'', $C_2(t)$, such that $A' = 2.5883$, $B' = 0.3125$, and $m' =4.3281$: 
for $C_1(t)$, $\textrm{mean}(E_r, k\geq 1) =-0.0618$, $\textrm{mean}(E_r, k\geq 4) =-0.0059$, $\max(|E_a|, k=30)$ = 1406;
for $C_2(t)$, $\textrm{mean}(E_r, k\geq 1) =0.0042$, $\textrm{mean}(E_r, k\geq 4) =-0.0098$, $\max(|E_a|, k=30)$ = 1709}
\label{fig:16bis}
\end{figure}

Such a numerical explosion of the parameters for low values of $m'$, therefore well illustrates the difficulty of the model $C_2(t)$ to naturally capture the concave growths. 

On the other hand, for a convex growth such as the one for period 1, the model $C_2(t)$, with $m'>1$, is absolutely representative of the evolution of official data corresponding to this period.
Indeed, figure \ref{fig:16bis} seems to show that the model $C_2(t)$ is, for a convexity, as representative as the model $C_1(t)$. More precisely, with respect to the latter, it offers a comparable precision to say the least, since its precision is better at the start (on three days) and similar beyond, as shown, with more nuance, by the numerical errors associated with figure \ref{fig:16bis}. 

Appropriated to the representation of the complexity that involves high numbers of cases, the model in $t^m$ does not have the vocation to represent small numbers of cases inherent to the short times: in other words, in reality, where the non-integer ope\-rator has only a sense at medium frequencies, the power-law dynamics, $t^m$, has only a sense at medium times, therefore beyond the short times. Besides, in the real case of the epidemic start, the nil slope of the initial behavior of $t^m$, is not stranger to the representativity lack of the FPM model at the short times.

 As for the precision of the FPM model in relation to the number of cases, it can be illustrated with the confirmed cases of the COVID-19 in Switzerland, given that these ones have been provided from a first case, recorded on February 25 2020. Between week 1 starting at this date with one confirmed case, and week 10 starting at April 28 2020 with 29264 confirmed cases, the maximum of the relative errors goes from 0.816 to 0.000976, by taking the intermediate values 0.321 and 0.0164 for weeks 2 and 5. That is to say that the precision is clearly improved by the case number (therefore by complexity).

\subsection{Rewriting of Verhulst model under the form of a model in $m^t$  whose properties are presented}
 The Verhulst model \eqref{eq:Verhulst} admits an immediate rewriting:
 \[
 f(t) = \frac{K}{1+\frac{K-a_0}{a_0} e^{-at}},
 \]
 or, by multiplying the numerator and the denominator by $a_0/(K-a_0)$:
 \[
 f(t) = \frac{\frac{Ka_0}{K-a_0}}{\frac{a_0}{K-a_0}+\left(e^{-a} \right) ^t},
 \]
 namely, by writing
 \[A=\frac{K a_0}{K-a_0}, \, B=\frac{a_0}{K-a_0} \quad \text{and}\quad m= e^{-a}:\]
 \[
 f(t) = \frac{A}{B+m^t},
 \]
 a model in $m^t$, with three parameters, $A$, $B$ and $m$, dual to the model in $t^m$, $A+Bt^m$.

 Used as an identification model, the function 
 \[
 f(t) = \frac{A}{B+m^t},
 \]
 is applied in a time interval such that $1\leq t\leq n $.
 
 This function is a particular solution of the differential equation
 \[
 f'(t) = -\ln(m) \left ( f(t) - B f^2(t) /A\right), \, f(1) = A/(B+m).
 \]
 
Its derivative is 
\[f'(t) = \frac{-A\ln (m) m^t}{\left (B+m^t \right)^2}.
\]

The function $f(t)$ is increasing if $A>0$ and $0<m<1$, or if $A<0$ and $m>1$, and decreasing otherwise.

Its second derivative is
\[
f''(t) =\displaystyle\frac{A \ln(m) m^t \left(m^t - B \right) }{\left (B+m^t \right)^3}.
\]
 
$f(t)$  can have an inflexion point in $t_0$ if $f''(t_0)=0$. This is only possible if $t_0 = \frac{\ln B}{\ln m}$ and if, moreover, $1<t_0<n$. 

Such a $t_0$ exists if $0<B<1$ and $0<m<1$, or if $B>1$ and $m>1$.

If $0<B<1$ and $0<m<1$, then $1<t_0<n$ if $m^n<B<m$.

If $B>1$ and $m>1$, then $1<t_0<n$ if $m<B<m^n$.

If $t_0$ is not between $1$ and $n$, then $f(t)$ is concave or convex if $t$ varies from 1 to $n$. 

If $t_0$ is between 1 and $n$, then the curve of $f(t)$ is a sigmoid, with two convex and concave parts. 


\subsection{Application of the model in $m^t$}
%

The model in $m^t $ is now applied, as an identification model, to period 1 which is the beginning of the epidemic in France. This application leads to the following numerical values: 
\[
A = 107.9402; \, B = 0.0022803;\, m= 0.7642119.
\]

Whereas period 1 presents a convexity, the value of $B$ belongs to an interval corresponding to an inflexion point, thus expressing that the model in $m^t$ is here inappropriate, as figure \ref{fig:Comp_tm_mt} shows it.

\begin{figure}[h]
\hspace{-0.6cm}
 \psfragfig[width=1.15\columnwidth]{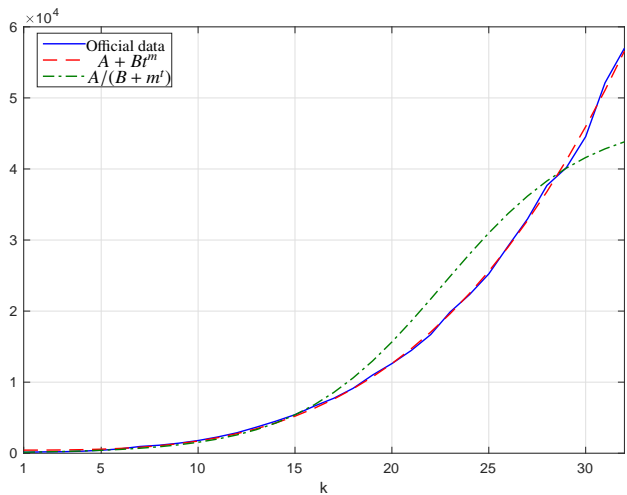}{
  \psfrag{cccccccccc}[c][][0.7]{\hspace{0.2cm} $A/(B+m^t)$}
    \psfrag{bbbbbbbbbb}[c][][0.7]{\hspace{0cm} $A+Bt^m$}
   \psfrag{aaaaaaaaaa}[c][][0.6]{\hspace{-0.1cm} Official data}
    }
\caption{
Comparative evolution, during period 1, of the contaminated recorded by the Ministry and obtained with the models in $t^m$ and $m^t$: the comparison well illustrates the representativity difference of the models}
\label{fig:Comp_tm_mt}
\end{figure}

In conclusion, figure \ref{fig:Comp_tm_mt} conveys an eloquent result: the comparative illustration of the representativity of the models in $t^m$ and $m^t$, clearly proves to be in favor of the model in $t^m$. But this drawback of the  model in $m^t$ is due to its strategy requirement, in the sense that  this model aims, with a unique parameterization, to represent both convex and concave phases of an epidemic growth. }}

\section{Comparative tables of the official data and the predictions obtained with the FPM model}\label{App:AnnexeE}

{

\begin{table*}[h]
\begin{center}
\captionof{table}{Basic period used for identification of the contaminated in lockdown phase in 2020} \label{tb:Ident2020}
\begin{tabular*}{\textwidth}{@{\extracolsep{\fill}} c|c|c|c|c|c|c|c|c|c}
\toprule
date &29/03	&30/03	&31/03	&01/04	&02/04	&03/04	&04/04	&05/04	& 06/04\\
\hline
$b_k$&39610	&46115	&51404	&56032	&60222	&64094	&67720	&71148	&74412\\
\hline
$a_k$&40174	&44550	&52128	&56989	&59105	&64338	&68605	&70478	&74390\\
\hline
$E_r$&0.0142  &$ -0.0339 $&   0.0141  &  0.0171& $  -0.0185$&    0.0038 &   0.0131 &$  -0.0094$&   -0.0003\\
\bottomrule	
\end{tabular*}
\end{center}
\end{table*}

\begin{table*}[h]
\begin{center}
\captionof{table}{Periods used for prediction of the contaminated in lockdown phase in 2020: $\textrm{mean}(E_r)=0.03568$ for week 1; $\textrm{mean}(E_r)=0.05525$ for week 2; $\textrm{mean}(E_r)=0.023988$ for week 3; $\textrm{mean}(E_r)=-0.023308$ for week 4; $\textrm{mean}(E_r)=-0.056094$ for week 5} \label{tb:Prediction2020}
\begin{tabular}{ c||c|c|c|c|c|c|c|c}
\toprule
\multirow{4}{*}{Week 1} &date	&07/04	&08/04	&09/04	&10/04	&11/04	&12/04	&13/04\\
\cline{2-9}\cline{2-9}
&$b_k$&77536	&80540	&83439	&86244	&88967	&91613	&94192\\
\cline{2-9}
&$a_k$& 78167	& 82048	& 86334	& 90676	&93790	& 95403	& 98076\\
\cline{2-9}
&$E_r$ & 0.0081   & 0.0187    &0.0347   & 0.0514    &0.0542    &0.0414   & 0.0412\\
\toprule	
\multirow{4}{*}{Week 2} &date	&14/04	&15/04	&16/04	& 17/04	& 18/04	& 19/04	& 20/04\\
\cline{2-9}\cline{2-9}
&$b_k$&96708	&99167	&101573	&103930	&106241	&108509	&110737\\
\cline{2-9}
&$a_k$& 103573	& 106206	& 108847	& 109252	& 111821	& 112206	& 114657\\
\cline{2-9}
&$E_r$&0.0710  &  0.0710   & 0.0716  &  0.0512  &  0.0525   & 0.0341   & 0.0354\\
\toprule
\multirow{4}{*}{Week 3} &date	&21/04	&22/04	&23/04	&24/04	&25/04	&26/04	&27/04\\
\cline{2-9}\cline{2-9}
&$b_k$&112928	&115083	&117205	&119295	&121354	&123386	&125389\\
\cline{2-9}
&$a_k$&117324	&119151	&120804&	122577	&124114	&124575	&125770\\
\cline{2-9}
&$E_r$&0.0389   & 0.0353&    0.0307&    0.0275&    0.0227&    0.0096&    0.0030\\
\toprule
\multirow{4}{*}{Week 4} &date	&28/04	&29/04	&30/04	&01/05	&02/05	&03/05	&04/05\\
\cline{2-9}\cline{2-9}
&$b_k$&127367	&129320	&131250	&133156	&135041	&136904	&138748\\
\cline{2-9}
&$a_k$&127290	&128442	&129581	&130185	&130979	&131287	&131863\\
\cline{2-9}
&$E_r$& $-0.0006$ & $   -0.0068$ &   $-0.0127$&$   -0.0223$&$   -0.0301$&$   -0.0410$&$   -0.0496$\\
\toprule
\multirow{4}{*}{Week 5} &date	&05/05	&06/05	&07/05	&08/05	&09/05	&10/05	&11/05\\
\cline{2-9}\cline{2-9}
&$b_k$&140572	&142377	&144165	&145934	&147687	&149424	&151145\\
\cline{2-9}
&$a_k$&132967	&137150	&137779	&138421	&138854	&139063	&139519\\
\cline{2-9}
&$E_r$& $ -0.0541$&$   -0.0367$&$   -0.0443$&$   -0.0515$&$   -0.0598$&$   -0.0693$&$   -0.0769$\\
\bottomrule	
\end{tabular}
\end{center}
\end{table*}


\begin{table*}[h]
\begin{center}
\captionof{table}{Basic period used for identification of the contaminated  in vaccination phase in 2021} \label{tb:Ident2021}
\begin{tabular*}{\textwidth}{@{\extracolsep{\fill}} c|c|c|c|c|c|c|c}
\toprule
date &26/04	&27/04	&28/04	&29/04	&30/04	&01/05	&02/05\\
\hline
$b_k$&5502175	&5537601	&5566631	&5592158	&5615361	&5636865	&5657053	\\
\hline
$a_k$&5503996	&5534313&5565852	&5592390	&5616689	&5642359	&5652247	\\
\hline
$E_r$&$0.00033$ &	$-0.00059$	& $- 0.00014$	&$0.000041$&$0.00024	$&$0.00097$&$-0.00085$\\
\bottomrule	
\end{tabular*}
\end{center}
\end{table*}

\begin{table*}[h]
\begin{center}
\captionof{table}{Periods used for prediction of the contaminated  in vaccination phase in 2021: $\textrm{mean}(E_r)= -0.000893 $ for week 1; $\textrm{mean}(E_r)=0.00017$ for week 2; $\textrm{mean}(E_r)=0.0010$ for week 3; $\textrm{mean}(E_r)= -0.00131$ for week 4; $\textrm{mean}(E_r)=-0.00430$ for week 5;$\textrm{mean}(E_r)=-0.00944$ for week 6} \label{tb:Prediction2021}
\begin{tabular}{ c||c|c|c|c|c|c|c|c}
\toprule
\multirow{4}{*}{Week 1} &date	&03/05	&04/05	&05/05	&06/05	&07/05	&08/05	&09/05\\
\cline{2-9}\cline{2-9}
&$b_k$&5676178	&5694419	&5711910	&5728753	&5745027	&5760798	&5776117\\
\cline{2-9}
&$a_k$&5656007	&5680378	&5706378	&5728090	&5747214	&5767959	&5777087\\
\cline{2-9}
&$E_r$& $ -0.0036$&$   -0.0025$&$   -0.0010$&$   -0.0001$&$    0.0004$&    0.0012&    0.0002\\
\toprule	
\multirow{4}{*}{Week 2} &date	&10/05	&11/05	&12/05	&13/05	&14/05	&15/05	&16/05\\
\cline{2-9}\cline{2-9}
&$b_k$&5791028	&5805569	&5819771	&5833661	&5847263	&5860598	&5873684\\
\cline{2-9}
&$a_k$&5780379	&5800170	&5821668&	5841129	&5848154	&5863839	&5877787\\
\cline{2-9}
&$E_r$& $-0.0018$&$    -0.0009$&    0.0003&    0.0013&    0.0002&    0.0006&    0.0007\\
\toprule
\multirow{4}{*}{Week 3} &date	&17/05	&18/05	&19/05	&20/05	&21/05	&22/05	&23/05\\
\cline{2-9}\cline{2-9}
&$b_k$&5886537	&5899171	&5911600	&5923835	&5935886	&5947764	&5959478\\
\cline{2-9}
&$a_k$&5881137	&5898347	&5917397	&5932812	&5945612	&5958223	&5967927\\
\cline{2-9}
&$E_r$& $-0.0009$&$   -0.0001$&    0.0010 &   0.0015   & 0.0016  &  0.0018  &  0.0014\\
\toprule
\multirow{4}{*}{Week 4} &date	&24/05	&25/05	&26/05	&27/05	&28/05	&29/05	&30/05\\
\cline{2-9}\cline{2-9}
&$b_k$&5971034	&5982442	&5993707	&6004835	&6015833	&6026707	&6037460\\
\cline{2-9}
&$a_k$&5970156	&5973311	&5985957	&5999890	&6011158	&6021833	&6030374\\
\cline{2-9}
&$E_r$& $-0.0001$&$   -0.0015$&$   -0.0013$&$   -0.0008$&$   -0.0008$&$   -0.0008 $&$   -0.0012$\\
\toprule
\multirow{4}{*}{Week 5} &date	&31/05	&01/06	&02/06	&03/06	&04/06	&05/06	&06/06\\
\cline{2-9}\cline{2-9}
&$b_k$&6048098	&6058625	&6069046	&6079363	&6089582	&6099704	&6109734\\
\cline{2-9}
&$a_k$&6031585	&6041433	&6050176	&6058340	&6065293	&6071947	&6077017\\
\cline{2-9}
&$E_r$& $-0.0027$&$   -0.0028$&$   -0.0031$&$   -0.0035$&$   -0.0040$&$   -0.0046$&$   -0.0054$\\
\toprule
\multirow{4}{*}{Week 6} &date	&07/06	&08/06	&09/06	&10/06	&11/06	&12/06	&13/06\\
\cline{2-9}\cline{2-9}
&$b_k$&6119674	&6129527	&6139297	&6148984	&6158593	&6168125	&6177582\\
\cline{2-9}
&$a_k$&6078181	&6084199	&6089756	&6094231	&6098102	&6102074	&6104929\\
\cline{2-9}
&$E_r$& -$0.0068$&$   -0.0074$&$   -0.0081$&$   -0.0089$&$   -0.0098$&$   -0.0107$&$   -0.0118$\\
\bottomrule	
\end{tabular}
\end{center}
\end{table*}


\begin{table*}[h]
\begin{center}
\captionof{table}{Basic period used for identification of the vaccinated from May 8 to May 14 2021} \label{tb:IdentVaccines2021}
\begin{tabular*}{\textwidth}{@{\extracolsep{\fill}} c|c|c|c|c|c|c|c}
\toprule
date &08/05	&09/05	&10/05	&11/05	&12/05	&13/05	&14/05\\
\hline
$b_k$&8257962	&8464151	&8670293	&8876403	&9082490	&9288558	&9494611	\\
\hline
$a_k$&8316714	&8371235	&8623700	&8917609	&9183962	&9260435	&9460813	\\
\hline
$E_r$&0.0071 &$  -0.0110$&$   -0.0054$&    0.0046&    0.0112&$   -0.0030$&$   -0.0036$\\
\bottomrule	
\end{tabular*}
\end{center}
\end{table*}

\begin{table*}[h]
\begin{center}
\captionof{table}{Periods used for prediction  of the vaccinated from May 15 to June 11 2021: $\textrm{mean}(E_r)=-0.02898$ for week 1; $\textrm{mean}(E_r)=-0.05833$ for week 2; $\textrm{mean}(E_r)=-0.04167$ for week 3; $\textrm{mean}(E_r)=-0.01112 $ for week 4} \label{tb:PredictionVaccines2021}
\begin{tabular}{ c||c|c|c|c|c|c|c|c}
\toprule
\multirow{4}{*}{Week 1} &date	&15/05	&16/05	&17/05	&18/05	&19/05	&20/05	&21/05\\
\cline{2-9}\cline{2-9}
&$b_k$&9700651	&9906680	&10112698	&10318707	&10524708	&10524708	&10936687\\
\cline{2-9}
&$a_k$&9579057	&9612256&	9789129	&9978383	&10158012	&10359164	&10577373\\
\cline{2-9}
&$E_r$&$ -0.0125$&$   -0.0297$&$   -0.0320$&$   -0.0330$&$   -0.0348$&$   -0.0157$&$   -0.0329$\\
\toprule	
\multirow{4}{*}{Week 2} &date	&22/05	&23/05	&24/05	&25/05	&26/05	&27/05	&28/05\\
\cline{2-9}\cline{2-9}
&$b_k$&11142666	&11348640	&11554607	&11760569	&11966527	&12172479	&12378426\\
\cline{2-9}
&$a_k$&10683925	&10709922	&10761760	&10985380	&11208870	&11472324	&11787631\\
\cline{2-9}
&$E_r$&$-0.0412$&$   -0.0563$&$   -0.0686$&$   -0.0659$&$   -0.0633$&$   -0.0575$&$   -0.0477$\\
\toprule
\multirow{4}{*}{Week 3} &date	&29/05	&30/05	&31/05	&01/06	&02/06	&03/06	&04/06\\
\cline{2-9}\cline{2-9}
&$b_k$&12584370	&12790309	&12996244	&13202175	&13408103	&13614027	&13819948\\
\cline{2-9}
&$a_k$&11968407	&12009982	&12249961	&12550529	&12825866	&13136728	&13496370\\
\cline{2-9}
&$E_r$&$-0.0489$&$   -0.0610$&$   -0.0574$&$   -0.0494$&$   -0.0434$&$   -0.0351$&$   -0.0234$\\
\toprule
\multirow{4}{*}{Week 4} &date	&05/06	&06/06	&07/06	&08/06	&09/06	&10/06	&11/06\\
\cline{2-9}\cline{2-9}
&$b_k$&14025865	&14231780	&14437691	&14643599	&14849505	&15055407	&15261307\\
\cline{2-9}
&$a_k$&13717098	&13777590	&14075211	&14411546	&14728908	&15064328	&15469804\\
\cline{2-9}
&$E_r$&$-0.0220$&$   -0.0319$&$   -0.0251$&$   -0.0158$&$   -0.0081$&    0.0006&    0.0137\\
\bottomrule	
\end{tabular}
\end{center}
\end{table*}

}





\begin{thebibliography}{64}
\expandafter\ifx\csname natexlab\endcsname\relax\def\natexlab#1{#1}\fi
\providecommand{\url}[1]{\texttt{#1}}
\providecommand{\href}[2]{#2}
\providecommand{\path}[1]{#1}
\providecommand{\DOIprefix}{doi:}
\providecommand{\ArXivprefix}{arXiv:}
\providecommand{\URLprefix}{URL: }
\providecommand{\Pubmedprefix}{pmid:}
\providecommand{\doi}[1]{\href{http://dx.doi.org/#1}{\path{#1}}}
\providecommand{\Pubmed}[1]{\href{pmid:#1}{\path{#1}}}
\providecommand{\bibinfo}[2]{#2}
\ifx\xfnm\relax \def\xfnm[#1]{\unskip,\space#1}\fi
\bibitem[{Andreatta et~al.(2005)Andreatta, P{\'e}rez~Lustres,
  P{\'e}rez~Lustres, Kovalenko, Ernsting, Murphy, Coleman and
  Berg}]{Andreatta2005}
\bibinfo{author}{Andreatta, D.}, \bibinfo{author}{P{\'e}rez~Lustres, J.L.},
  \bibinfo{author}{P{\'e}rez~Lustres, J.L.}, \bibinfo{author}{Kovalenko, S.A.},
  \bibinfo{author}{Ernsting, N.P.}, \bibinfo{author}{Murphy, C.J.},
  \bibinfo{author}{Coleman, R.S.}, \bibinfo{author}{Berg, M.A.},
  \bibinfo{year}{2005}.
\newblock \bibinfo{title}{Power-law solvation dynamics in dna over six decades
  in time}.
\newblock \bibinfo{journal}{Journal of the American Chemical Society}
  \bibinfo{volume}{127}, \bibinfo{pages}{7270---7271}.
\newblock \DOIprefix\doi{10.1021/ja044177v}.
\bibitem[{Arfan et~al.(2020)Arfan, Shah, Abdeljawad, Mlaiki and
  Ullah}]{Arfan2020}
\bibinfo{author}{Arfan, M.}, \bibinfo{author}{Shah, K.},
  \bibinfo{author}{Abdeljawad, T.}, \bibinfo{author}{Mlaiki, N.},
  \bibinfo{author}{Ullah, A.}, \bibinfo{year}{2020}.
\newblock \bibinfo{title}{A caputo power law model predicting the spread of the
  covid-19 outbreak in pakistan}.
\newblock \bibinfo{journal}{Alexandria Engineering Journal}
  \DOIprefix\doi{https://doi.org/10.1016/j.aej.2020.09.011}.
\bibitem[{Atangana(2020)}]{Atangana2020}
\bibinfo{author}{Atangana, A.}, \bibinfo{year}{2020}.
\newblock \bibinfo{title}{Modelling the spread of covid-19 with new
  fractal-fractional operators: Can the lockdown save mankind before
  vaccination?}
\newblock \bibinfo{journal}{Chaos, Solitons \& Fractals} \bibinfo{volume}{136},
  \bibinfo{pages}{109860}.
\newblock \DOIprefix\doi{https://doi.org/10.1016/j.chaos.2020.109860}.
\bibitem[{Battaglia et~al.(2000)Battaglia, Le~Lay, Batsale, Oustaloup and
  Cois}]{Battaglia2000}
\bibinfo{author}{Battaglia, J.L.}, \bibinfo{author}{Le~Lay, L.},
  \bibinfo{author}{Batsale, J.C.}, \bibinfo{author}{Oustaloup, A.},
  \bibinfo{author}{Cois, O.}, \bibinfo{year}{2000}.
\newblock \bibinfo{title}{Heat flux estimation through inverted non integer
  identification models}.
\newblock \bibinfo{journal}{International Journal of Thermal Science}
  \bibinfo{volume}{39}, \bibinfo{pages}{374--389}.
\newblock \DOIprefix\doi{10.1016/S1290-0729(00)00220-9}.
\bibitem[{Carli et~al.(2020)Carli, Cavone, Epicoco, Scarabaggio and
  Dotoli}]{Carli2020}
\bibinfo{author}{Carli, R.}, \bibinfo{author}{Cavone, G.},
  \bibinfo{author}{Epicoco, N.}, \bibinfo{author}{Scarabaggio, P.},
  \bibinfo{author}{Dotoli, M.}, \bibinfo{year}{2020}.
\newblock \bibinfo{title}{Model predictive control to mitigate the covid-19
  outbreak in a multi-region scenario}.
\newblock \bibinfo{journal}{Annual Reviews in Control} \bibinfo{volume}{50},
  \bibinfo{pages}{373--393}.
\newblock \DOIprefix\doi{10.1016/j.arcontrol.2020.09.005}.
\bibitem[{Chen et~al.(2020)Chen, Rui, Wang, Zhao, Cui and Yin}]{Chen2020}
\bibinfo{author}{Chen, T.}, \bibinfo{author}{Rui, J.}, \bibinfo{author}{Wang,
  Q.}, \bibinfo{author}{Zhao, Z.}, \bibinfo{author}{Cui, J.A.},
  \bibinfo{author}{Yin, L.}, \bibinfo{year}{2020}.
\newblock \bibinfo{title}{A mathematical model for simulating the transmission
  of wuhan novel coronavirus}.
\newblock \bibinfo{journal}{bioRxiv} \DOIprefix\doi{10.1101/2020.01.19.911669}.
\bibitem[{De-Leon and Pederiva(2020)}]{De-Leon2020}
\bibinfo{author}{De-Leon, H.}, \bibinfo{author}{Pederiva, F.},
  \bibinfo{year}{2020}.
\newblock \bibinfo{title}{Particle modeling of the spreading of coronavirus
  disease (covid-19)}.
\newblock \bibinfo{journal}{Physics of Fluids} \bibinfo{volume}{32},
  \bibinfo{pages}{087113}.
\newblock \DOIprefix\doi{10.1063/5.0020565}.
\bibitem[{De~Visscher(2020)}]{De-Visscher2020}
\bibinfo{author}{De~Visscher, A.}, \bibinfo{year}{2020}.
\newblock \bibinfo{title}{The covid-19 pandemic: model-based evaluation of
  non-pharmaceutical interventions and prognoses}.
\newblock \bibinfo{journal}{Nonlinear Dynamics} \bibinfo{volume}{101},
  \bibinfo{pages}{1871--1887}.
\newblock \DOIprefix\doi{10.1007/s11071-020-05861-7}.
\bibitem[{{Dell'Anna}(2020)}]{DellAnna2020}
\bibinfo{author}{{Dell'Anna}, L.}, \bibinfo{year}{2020}.
\newblock \bibinfo{title}{Solvable delay model for epidemic spreading: the case
  of covid-19 in italy}.
\newblock \bibinfo{journal}{Scientific Reports} \bibinfo{volume}{10},
  \bibinfo{pages}{15763}.
\newblock \DOIprefix\doi{10.1038/s41598-020-72529-y}.
\bibitem[{Demongeot et~al.(2020)Demongeot, Griette and Magal}]{Demongeot2020}
\bibinfo{author}{Demongeot, J.}, \bibinfo{author}{Griette, Q.},
  \bibinfo{author}{Magal, P.}, \bibinfo{year}{2020}.
\newblock \bibinfo{title}{Si epidemic model applied to covid-19 data in
  mainland china}.
\newblock \bibinfo{journal}{medRxiv}
  \DOIprefix\doi{10.1101/2020.10.19.20214528}.
\bibitem[{Duan et~al.(2015)Duan, Fan, Zhang, Guo and Qiu}]{Duan2015}
\bibinfo{author}{Duan, W.}, \bibinfo{author}{Fan, Z.}, \bibinfo{author}{Zhang,
  P.}, \bibinfo{author}{Guo, G.}, \bibinfo{author}{Qiu, X.},
  \bibinfo{year}{2015}.
\newblock \bibinfo{title}{Mathematical and computational approaches to epidemic
  modeling: a comprehensive review}.
\newblock \bibinfo{journal}{Frontiers of Computer Science} \bibinfo{volume}{9},
  \bibinfo{pages}{806--826}.
\newblock \DOIprefix\doi{10.1007/s11704-014-3369-2}.
\bibitem[{Efimov and Ushirobira(2021)}]{Efimov2021}
\bibinfo{author}{Efimov, D.}, \bibinfo{author}{Ushirobira, R.},
  \bibinfo{year}{2021}.
\newblock \bibinfo{title}{On an interval prediction of covid-19 development
  based on a seir epidemic model}.
\newblock \bibinfo{journal}{Annual Reviews in Control}
  \DOIprefix\doi{10.1016/j.arcontrol.2021.01.006}.
\bibitem[{Erd\'{e}lyi(1962)}]{Erdelyi1962}
\bibinfo{author}{Erd\'{e}lyi, A.}, \bibinfo{year}{1962}.
\newblock \bibinfo{title}{Operational Calculus and Generalized Functions}.
\newblock \bibinfo{publisher}{Holt, Rinehart and Winston},
  \bibinfo{address}{New-York}.
\bibitem[{Ferguson et~al.(2005)Ferguson, Cummings, Cauchemez, Fraser, Riley,
  Meeyai, Iamsirithaworn and Burke}]{Ferguson2005}
\bibinfo{author}{Ferguson, N.M.}, \bibinfo{author}{Cummings, D.A.T.},
  \bibinfo{author}{Cauchemez, S.}, \bibinfo{author}{Fraser, C.},
  \bibinfo{author}{Riley, S.}, \bibinfo{author}{Meeyai, A.},
  \bibinfo{author}{Iamsirithaworn, S.}, \bibinfo{author}{Burke, D.S.},
  \bibinfo{year}{2005}.
\newblock \bibinfo{title}{Strategies for containing an emerging influenza
  pandemic in southeast asia}.
\newblock \bibinfo{journal}{Nature} \bibinfo{volume}{437},
  \bibinfo{pages}{209--214}.
\newblock \DOIprefix\doi{10.1038/nature04017}.
\bibitem[{Frieman et~al.(2008)Frieman, Turner and Huterer}]{Frieman2008}
\bibinfo{author}{Frieman, J.A.}, \bibinfo{author}{Turner, M.S.},
  \bibinfo{author}{Huterer, D.}, \bibinfo{year}{2008}.
\newblock \bibinfo{title}{Dark energy and the accelerating universe}.
\newblock \bibinfo{journal}{Annual Review of Astronomy and Astrophysics}
  \bibinfo{volume}{46}, \bibinfo{pages}{385--432}.
\newblock \DOIprefix\doi{10.1146/annurev.astro.46.060407.145243}.
\bibitem[{Gabaix et~al.(2003)Gabaix, Gopikrishnan, Plerou and
  Stanley}]{Gabaix2003}
\bibinfo{author}{Gabaix, X.}, \bibinfo{author}{Gopikrishnan, P.},
  \bibinfo{author}{Plerou, V.}, \bibinfo{author}{Stanley, H.},
  \bibinfo{year}{2003}.
\newblock \bibinfo{title}{A theory of power-law distributions in financial
  market fluctuations}.
\newblock \bibinfo{journal}{Nature} \bibinfo{volume}{423},
  \bibinfo{pages}{267--270}.
\newblock \DOIprefix\doi{10.1038/nature01624}.
\bibitem[{Ganyani et~al.(2018)Ganyani, Roosa, Faes, Hens and
  Chowell}]{Ganyani2018}
\bibinfo{author}{Ganyani, T.}, \bibinfo{author}{Roosa, K.},
  \bibinfo{author}{Faes, C.}, \bibinfo{author}{Hens, N.},
  \bibinfo{author}{Chowell, G.}, \bibinfo{year}{2018}.
\newblock \bibinfo{title}{Assessing the relationship between epidemic growth
  scaling and epidemic size: The 2014-16 ebola epidemic in west africa}.
\newblock \bibinfo{journal}{Epidemiology and infection} \bibinfo{volume}{147},
  \bibinfo{pages}{1--6}.
\newblock \DOIprefix\doi{10.1017/S0950268818002819}.
\bibitem[{Garetto et~al.(2021)Garetto, Leonardi and Torrisi}]{Garetto2021}
\bibinfo{author}{Garetto, M.}, \bibinfo{author}{Leonardi, E.},
  \bibinfo{author}{Torrisi, G.}, \bibinfo{year}{2021}.
\newblock \bibinfo{title}{A time-modulated hawkes process to model the spread
  of covid-19 and the impact of countermeasures}.
\newblock \bibinfo{journal}{Annual Reviews in Control}
  \DOIprefix\doi{10.1016/j.arcontrol.2021.02.002}.
\bibitem[{Guan et~al.(2020)Guan, Prieur, Zhang, Prieur, Georges and
  Bellemain}]{Guan2020}
\bibinfo{author}{Guan, L.}, \bibinfo{author}{Prieur, C.},
  \bibinfo{author}{Zhang, L.}, \bibinfo{author}{Prieur, C.},
  \bibinfo{author}{Georges, D.}, \bibinfo{author}{Bellemain, P.},
  \bibinfo{year}{2020}.
\newblock \bibinfo{title}{Transport effect of covid-19 pandemic in france}, in:
  \bibinfo{booktitle}{{{J}ourn\'ees {N}ationales {A}utomtique de la {SAGIP} -
  {G}roupe de {T}ravail {I}dentification et COVID-19}},
  \bibinfo{address}{Lille, France}.
\bibitem[{Harman et~al.(2009)Harman, Sivapalan and Kumar}]{Harman2009}
\bibinfo{author}{Harman, C.J.}, \bibinfo{author}{Sivapalan, M.},
  \bibinfo{author}{Kumar, P.}, \bibinfo{year}{2009}.
\newblock \bibinfo{title}{Power law catchment-scale recessions arising from
  heterogeneous linear small-scale dynamics}.
\newblock \bibinfo{journal}{Water Resources Research} \bibinfo{volume}{45}.
\newblock \DOIprefix\doi{10.1029/2008WR007392}.
\bibitem[{He et~al.(2020)He, Peng and Sun}]{He2020}
\bibinfo{author}{He, S.}, \bibinfo{author}{Peng, Y.}, \bibinfo{author}{Sun,
  K.}, \bibinfo{year}{2020}.
\newblock \bibinfo{title}{Seir modeling of the covid-19 and its dynamics}.
\newblock \bibinfo{journal}{Nonlinear Dynamics} \bibinfo{volume}{101},
  \bibinfo{pages}{1667--1680}.
\newblock \DOIprefix\doi{10.1007/s11071-020-05743-y}.
\bibitem[{Huang and Qi(2020)}]{HuangJianzhe2020}
\bibinfo{author}{Huang, J.}, \bibinfo{author}{Qi, G.}, \bibinfo{year}{2020}.
\newblock \bibinfo{title}{Effects of control measures on the dynamics of
  covid-19 and double-peak behavior in spain}.
\newblock \bibinfo{journal}{Nonlinear Dynamics} \bibinfo{volume}{101},
  \bibinfo{pages}{1889--1899}.
\newblock \DOIprefix\doi{10.1007/s11071-020-05901-2}.
\bibitem[{Huang et~al.(2020)Huang, Qiao, Wang and Tung}]{Huang2020}
\bibinfo{author}{Huang, N.E.}, \bibinfo{author}{Qiao, F.},
  \bibinfo{author}{Wang, Q.}, \bibinfo{author}{Tung, K.K.},
  \bibinfo{year}{2020}.
\newblock \bibinfo{title}{Herd immunity vs suppressed equilibrium in covid-19
  pandemic: different goals require different models for tracking}.
\newblock \bibinfo{journal}{medRxiv}
  \DOIprefix\doi{10.1101/2020.03.28.20046177}.
\bibitem[{Iacus et~al.(2020)Iacus, Santamaria, Sermi, Spyratos, Tarchi and
  Vespe}]{Iacus2020}
\bibinfo{author}{Iacus, S.M.}, \bibinfo{author}{Santamaria, C.},
  \bibinfo{author}{Sermi, F.}, \bibinfo{author}{Spyratos, S.},
  \bibinfo{author}{Tarchi, D.}, \bibinfo{author}{Vespe, M.},
  \bibinfo{year}{2020}.
\newblock \bibinfo{title}{Human mobility and covid-19 initial dynamics}.
\newblock \bibinfo{journal}{Nonlinear Dynamics} \bibinfo{volume}{101},
  \bibinfo{pages}{1901--1919}.
\newblock \DOIprefix\doi{10.1007/s11071-020-05854-6}.
\bibitem[{Ivorra et~al.(2020)Ivorra, Ferr{\'a}ndez, Vela-P{\'e}rez and
  Ramos}]{Ivorra2020}
\bibinfo{author}{Ivorra, B.}, \bibinfo{author}{Ferr{\'a}ndez, M.},
  \bibinfo{author}{Vela-P{\'e}rez, M.}, \bibinfo{author}{Ramos, A.},
  \bibinfo{year}{2020}.
\newblock \bibinfo{title}{Mathematical modeling of the spread of the
  coronavirus disease 2019 (covid-19) taking into account the undetected
  infections. the case of china}.
\newblock \bibinfo{journal}{Communications in Nonlinear Science and Numerical
  Simulation} \bibinfo{volume}{88}, \bibinfo{pages}{105303}.
\newblock \DOIprefix\doi{https://doi.org/10.1016/j.cnsns.2020.105303}.
\bibitem[{Kermack and McKendrick(1927)}]{Kermack1927}
\bibinfo{author}{Kermack, W.}, \bibinfo{author}{McKendrick, A.},
  \bibinfo{year}{1927}.
\newblock \bibinfo{title}{A contribution to the mathematical theory of
  epidemics}.
\newblock \bibinfo{journal}{Proc. R. Soc. Lond} \bibinfo{volume}{A},
  \bibinfo{pages}{700--721}.
\newblock \DOIprefix\doi{http://doi.org/10.1098/rspa.1927.0118}.
\bibitem[{Korabel et~al.(2007)Korabel, Zaslavsky and Tarasov}]{Korabel2007}
\bibinfo{author}{Korabel, N.}, \bibinfo{author}{Zaslavsky, G.},
  \bibinfo{author}{Tarasov, V.}, \bibinfo{year}{2007}.
\newblock \bibinfo{title}{Coupled oscillators with power-law interaction and
  their fractional dynamics analogues}.
\newblock \bibinfo{journal}{Communications in Nonlinear Science and Numerical
  Simulation} \bibinfo{volume}{12}, \bibinfo{pages}{1405 -- 1417}.
\newblock \DOIprefix\doi{10.1016/j.cnsns.2006.03.015}.
\bibitem[{Kwuimy et~al.(2020)Kwuimy, Nazari, Jiao, Rohani and
  Nataraj}]{Kwuimy2020}
\bibinfo{author}{Kwuimy, C.A.K.}, \bibinfo{author}{Nazari, F.},
  \bibinfo{author}{Jiao, X.}, \bibinfo{author}{Rohani, P.},
  \bibinfo{author}{Nataraj, C.}, \bibinfo{year}{2020}.
\newblock \bibinfo{title}{Nonlinear dynamic analysis of an epidemiological
  model for covid-19 including public behavior and government action}.
\newblock \bibinfo{journal}{Nonlinear Dynamics} \bibinfo{volume}{101},
  \bibinfo{pages}{1545--1559}.
\newblock \DOIprefix\doi{10.1007/s11071-020-05815-z}.
\bibitem[{Kyurkchiev et~al.(2020)Kyurkchiev, Iliev and Rahnev}]{Kyurkchiev2020}
\bibinfo{author}{Kyurkchiev, N.}, \bibinfo{author}{Iliev, A.},
  \bibinfo{author}{Rahnev, A.}, \bibinfo{year}{2020}.
\newblock \bibinfo{title}{On the verhulst growth model with polynomial variable
  transfer. some applications}.
\newblock \bibinfo{journal}{International. journal of differential, equations
  and applications} \bibinfo{volume}{19}, \bibinfo{pages}{15--32}.
\bibitem[{{Le M\'ehaut\'e}(2021)}]{LeMehaute2021}
\bibinfo{author}{{Le M\'ehaut\'e}, A.}, \bibinfo{year}{2021}.
\newblock \bibinfo{title}{About the theoretical roots for managing the third
  middle illustration using covid-19 dynamics in france}.
\newblock \bibinfo{journal}{Noworczesne Syslemy Zazadzania. Kwartalny raport
  naukowy} \bibinfo{volume}{To appear}.
\bibitem[{Li et~al.(2020)Li, Zhou and Lu}]{Li2020}
\bibinfo{author}{Li, W.}, \bibinfo{author}{Zhou, J.}, \bibinfo{author}{Lu,
  J.a.}, \bibinfo{year}{2020}.
\newblock \bibinfo{title}{The effect of behavior of wearing masks on epidemic
  dynamics}.
\newblock \bibinfo{journal}{Nonlinear Dynamics} \bibinfo{volume}{101},
  \bibinfo{pages}{1995--2001}.
\newblock \DOIprefix\doi{10.1007/s11071-020-05759-4}.
\bibitem[{Liang et~al.(2009)Liang, Wang, Lee, Peng and Su}]{Liang2009}
\bibinfo{author}{Liang, Z.P.}, \bibinfo{author}{Wang, X.D.},
  \bibinfo{author}{Lee, D.J.}, \bibinfo{author}{Peng, X.F.},
  \bibinfo{author}{Su, A.}, \bibinfo{year}{2009}.
\newblock \bibinfo{title}{Spreading dynamics of power-law fluid droplets}.
\newblock \bibinfo{journal}{Journal of Physics: Condensed Matter}
  \bibinfo{volume}{21}, \bibinfo{pages}{464117}.
\newblock \DOIprefix\doi{10.1088/0953-8984/21/46/464117}.
\bibitem[{Liouville(1832)}]{Liouville1832}
\bibinfo{author}{Liouville, J.}, \bibinfo{year}{1832}.
\newblock \bibinfo{title}{M\'{e}moire sur quelques questions de
  g\'{e}om\'{e}trie et de m\'{e}canique et sur un nouveau genre de calcul pour
  r\'{e}soudre ces \'{e}quations}.
\newblock \bibinfo{journal}{Journal de l'{E}cole {P}olytechnique}
  \bibinfo{volume}{13}, \bibinfo{pages}{71--162}.
\bibitem[{Liu et~al.(2020)Liu, Zheng and Balachandran}]{Liu2020}
\bibinfo{author}{Liu, X.}, \bibinfo{author}{Zheng, X.},
  \bibinfo{author}{Balachandran, B.}, \bibinfo{year}{2020}.
\newblock \bibinfo{title}{Covid-19: data-driven dynamics, statistical and
  distributed delay models, and observations}.
\newblock \bibinfo{journal}{Nonlinear Dynamics} \bibinfo{volume}{101},
  \bibinfo{pages}{1527--1543}.
\newblock \DOIprefix\doi{10.1007/s11071-020-05863-5}.
\bibitem[{Lu et~al.(2020)Lu, Yu, Chen, Ren, Xu, Wang and Yin}]{Lu2020}
\bibinfo{author}{Lu, Z.}, \bibinfo{author}{Yu, Y.}, \bibinfo{author}{Chen, Y.},
  \bibinfo{author}{Ren, G.}, \bibinfo{author}{Xu, C.}, \bibinfo{author}{Wang,
  S.}, \bibinfo{author}{Yin, Z.}, \bibinfo{year}{2020}.
\newblock \bibinfo{title}{A fractional-order seihdr model for covid-19 with
  inter-city networked coupling effects}.
\newblock \bibinfo{journal}{Nonlinear Dynamics} \bibinfo{volume}{101},
  \bibinfo{pages}{1717--1730}.
\newblock \DOIprefix\doi{10.1007/s11071-020-05848-4}.
\bibitem[{Mendes and {da Silva}(2009)}]{Mendes2009}
\bibinfo{author}{Mendes, G.}, \bibinfo{author}{{da Silva}, L.},
  \bibinfo{year}{2009}.
\newblock \bibinfo{title}{Generating more realistic complex networks from
  power-law distribution of fitness generating more realistic complex networks
  from power-law distribution of fitness}.
\newblock \bibinfo{journal}{Braz. J. Phys.} \bibinfo{volume}{39}.
\newblock \DOIprefix\doi{10.1590/S0103-97332009000400013}.
\bibitem[{Meyer and Held(2014)}]{Meyer2014}
\bibinfo{author}{Meyer, S.}, \bibinfo{author}{Held, L.}, \bibinfo{year}{2014}.
\newblock \bibinfo{title}{Power-law models for infectious disease spread}.
\newblock \bibinfo{journal}{The Annals of Applied Statistics}
  \bibinfo{volume}{8}, \bibinfo{pages}{1612--1639}.
\newblock \DOIprefix\doi{10.1214/14-AOAS743}.
\bibitem[{Oustaloup(1995)}]{Oustaloup1995a}
\bibinfo{author}{Oustaloup, A.}, \bibinfo{year}{1995}.
\newblock \bibinfo{title}{La d\'{e}rivation non-enti\`{e}re : th\'{e}orie,
  synth\`{e}se et applications}.
\newblock \bibinfo{publisher}{Herm\`{e}s}, \bibinfo{address}{Paris}.
\bibitem[{Oustaloup(2014)}]{Oustaloup2014}
\bibinfo{author}{Oustaloup, A.}, \bibinfo{year}{2014}.
\newblock \bibinfo{title}{Diversity and Non-integer Differentiation for System
  Dynamics}.
\newblock \bibinfo{publisher}{Wiley-ISTE}.
\bibitem[{Oustaloup et~al.(2000)Oustaloup, Levron, Mathieu and
  Nanot}]{Oustaloup2000}
\bibinfo{author}{Oustaloup, A.}, \bibinfo{author}{Levron, F.},
  \bibinfo{author}{Mathieu, B.}, \bibinfo{author}{Nanot, F.},
  \bibinfo{year}{2000}.
\newblock \bibinfo{title}{Frequency-band complex noninteger differentiator:
  characterization and synthesis}.
\newblock \bibinfo{journal}{IEEE Transactions on Circuits and Systems I:
  Fundamental Theory and Applications} \bibinfo{volume}{47},
  \bibinfo{pages}{25--39}.
\newblock \DOIprefix\doi{10.1109/81.817385}.
\bibitem[{R\'emond and R\'emond(2020)}]{Remond2020}
\bibinfo{author}{R\'emond, J.}, \bibinfo{author}{R\'emond, Y.},
  \bibinfo{year}{2020}.
\newblock \bibinfo{title}{Sur une mod\'elisation simplifi\'ee de l'\'epid\'emie
  du covid-19 de 2020}.
\newblock \URLprefix \url{https://hal.archives-ouvertes.fr/hal-02551464}.
  \bibinfo{note}{working paper or preprint}.
\bibitem[{Richards(1959)}]{Richards1959}
\bibinfo{author}{Richards, F.}, \bibinfo{year}{1959}.
\newblock \bibinfo{title}{A flexible growth function for empirical use}.
\newblock \bibinfo{journal}{Journal of Experimental Botany}
  \bibinfo{volume}{10}, \bibinfo{pages}{290--300}.
\bibitem[{Rohith and Devika(2020)}]{Rohith2020}
\bibinfo{author}{Rohith, G.}, \bibinfo{author}{Devika, K.B.},
  \bibinfo{year}{2020}.
\newblock \bibinfo{title}{Dynamics and control of covid-19 pandemic with
  nonlinear incidence rates}.
\newblock \bibinfo{journal}{Nonlinear Dynamics} \bibinfo{volume}{101},
  \bibinfo{pages}{2013--2026}.
\newblock \DOIprefix\doi{10.1007/s11071-020-05774-5}.
\bibitem[{Roosa et~al.(2020)Roosa, Lee, Luo, Kirpich, Rothenberg, Hyman, Yan
  and Chowell}]{Roosa2020}
\bibinfo{author}{Roosa, K.}, \bibinfo{author}{Lee, Y.}, \bibinfo{author}{Luo,
  R.}, \bibinfo{author}{Kirpich, A.}, \bibinfo{author}{Rothenberg, R.},
  \bibinfo{author}{Hyman, J.M.}, \bibinfo{author}{Yan, P.},
  \bibinfo{author}{Chowell, G.}, \bibinfo{year}{2020}.
\newblock \bibinfo{title}{Short-term forecasts of the covid-19 epidemic in
  guangdong and zhejiang, china: February 13-23, 2020}.
\newblock \bibinfo{journal}{Journal of clinical medicine} \bibinfo{volume}{9},
  \bibinfo{pages}{596}.
\newblock \DOIprefix\doi{10.3390/jcm9020596}.
\bibitem[{Santamaria(2017)}]{Santamaria2017}
\bibinfo{author}{Santamaria, J.}, \bibinfo{year}{2017}.
\newblock \bibinfo{title}{De la simplicit{\'e} {\`a} la complexit{\'e}}.
\newblock {Le Monde}, \bibinfo{publisher}{Voyage dans le cosmos}.
\bibitem[{Satsuma et~al.(2004)Satsuma, Willox, Ramani, Grammaticos and
  Carstea}]{Satsuma2004}
\bibinfo{author}{Satsuma, J.}, \bibinfo{author}{Willox, R.},
  \bibinfo{author}{Ramani, A.}, \bibinfo{author}{Grammaticos, B.},
  \bibinfo{author}{Carstea, A.}, \bibinfo{year}{2004}.
\newblock \bibinfo{title}{Extending the sir epidemic model}.
\newblock \bibinfo{journal}{Physica A: Statistical Mechanics and its
  Applications} \bibinfo{volume}{336}, \bibinfo{pages}{369 -- 375}.
\newblock \DOIprefix\doi{https://doi.org/10.1016/j.physa.2003.12.035}.
\bibitem[{Savageau(1970)}]{Savageau1970}
\bibinfo{author}{Savageau, M.A.}, \bibinfo{year}{1970}.
\newblock \bibinfo{title}{Biochemical systems analysis: Iii. dynamic solutions
  using a power-law approximation}.
\newblock \bibinfo{journal}{Journal of Theoretical Biology}
  \bibinfo{volume}{26}, \bibinfo{pages}{215 -- 226}.
\newblock \DOIprefix\doi{10.1016/S0022-5193(70)80013-3}.
\bibitem[{Scharbarg et~al.(2020)Scharbarg, Moog, Mauduit and
  Califano}]{Scharbarg2020}
\bibinfo{author}{Scharbarg, E.}, \bibinfo{author}{Moog, C.H.},
  \bibinfo{author}{Mauduit, N.}, \bibinfo{author}{Califano, C.},
  \bibinfo{year}{2020}.
\newblock \bibinfo{title}{From the hospital scale to nationwide: observability
  and identification of models for the covid-19 epidemic waves}.
\newblock \bibinfo{journal}{Annual Reviews in Control} \bibinfo{volume}{50},
  \bibinfo{pages}{409--416}.
\newblock \DOIprefix\doi{10.1016/j.arcontrol.2020.09.007}.
\bibitem[{Sher et~al.(2008)Sher, Smith, Dalgarno, Warburton, Chen, Dobson,
  Daniels, Pickett and O'Brien}]{Sher2008}
\bibinfo{author}{Sher, P.H.}, \bibinfo{author}{Smith, J.M.},
  \bibinfo{author}{Dalgarno, P.A.}, \bibinfo{author}{Warburton, R.J.},
  \bibinfo{author}{Chen, X.}, \bibinfo{author}{Dobson, P.J.},
  \bibinfo{author}{Daniels, S.M.}, \bibinfo{author}{Pickett, N.L.},
  \bibinfo{author}{O'Brien, P.}, \bibinfo{year}{2008}.
\newblock \bibinfo{title}{Power law carrier dynamics in semiconductor
  nanocrystals at nanosecond timescales}.
\newblock \bibinfo{journal}{Applied Physics Letters} \bibinfo{volume}{92},
  \bibinfo{pages}{101111}.
\newblock \DOIprefix\doi{10.1063/1.2894193}.
\bibitem[{Tanner(1979)}]{Tanner1979}
\bibinfo{author}{Tanner, L.H.}, \bibinfo{year}{1979}.
\newblock \bibinfo{title}{The spreading of silicone oil drops on horizontal
  surfaces}.
\newblock \bibinfo{journal}{Journal of Physics D: Applied Physics}
  \bibinfo{volume}{12}, \bibinfo{pages}{1473--1484}.
\newblock \DOIprefix\doi{10.1088/0022-3727/12/9/009}.
\bibitem[{Tolle(2003)}]{Tolle2003}
\bibinfo{author}{Tolle, J.}, \bibinfo{year}{2003}.
\newblock \bibinfo{title}{Can growth be faster than exponential, and just how
  slow is the logarithm?}
\newblock \bibinfo{journal}{The Mathematical Gazette} \bibinfo{volume}{87},
  \bibinfo{pages}{522--525}.
\newblock \DOIprefix\doi{10.1017/S0025557200173802}.
\bibitem[{Trigeassou and Maamri(2019)}]{Trigeassou2019b}
\bibinfo{author}{Trigeassou, J.}, \bibinfo{author}{Maamri, N.},
  \bibinfo{year}{2019}.
\newblock \bibinfo{title}{The Distributed Model of the Fractional Integrator}.
  \bibinfo{publisher}{John Wiley \& Sons, Ltd}. chapter~\bibinfo{chapter}{6}.
\newblock pp. \bibinfo{pages}{127--158}.
\newblock \DOIprefix\doi{10.1002/9781119476917.ch6}.
\bibitem[{Tuan et~al.(2020)Tuan, Mohammadi and Rezapour}]{Tuan2020}
\bibinfo{author}{Tuan, N.H.}, \bibinfo{author}{Mohammadi, H.},
  \bibinfo{author}{Rezapour, S.}, \bibinfo{year}{2020}.
\newblock \bibinfo{title}{A mathematical model for covid-19 transmission by
  using the caputo fractional derivative}.
\newblock \bibinfo{journal}{Chaos, Solitons \& Fractals} \bibinfo{volume}{140},
  \bibinfo{pages}{110107}.
\newblock \DOIprefix\doi{https://doi.org/10.1016/j.chaos.2020.110107}.
\bibitem[{Verhulst(1838)}]{Verhulst1838}
\bibinfo{author}{Verhulst, P.F.}, \bibinfo{year}{1838}.
\newblock \bibinfo{title}{Notice sur la loi que la population poursuit dans son
  accroissement}.
\newblock \bibinfo{journal}{Correspondance math\'ematique et physique} ,
  \bibinfo{pages}{113--121}.
\bibitem[{Verhulst(1845)}]{Verhulst1845}
\bibinfo{author}{Verhulst, P.F.}, \bibinfo{year}{1845}.
\newblock \bibinfo{title}{Recherches math\'ematiques sur la loi d'accroissement
  de la population}.
\newblock \bibinfo{journal}{Nouveaux m\'emoires de l'Acad\'emie Royale des
  Sciences et Belles-Lettres de Bruxelles} \bibinfo{volume}{18},
  \bibinfo{pages}{14--54}.
\bibitem[{Viboud et~al.(2016)Viboud, Simonsen and Chowell}]{Viboud2016}
\bibinfo{author}{Viboud, C.}, \bibinfo{author}{Simonsen, L.},
  \bibinfo{author}{Chowell, G.}, \bibinfo{year}{2016}.
\newblock \bibinfo{title}{A generalized-growth model to characterize the early
  ascending phase of infectious disease outbreaks}.
\newblock \bibinfo{journal}{Epidemics} \bibinfo{volume}{15}, \bibinfo{pages}{27
  -- 37}.
\newblock \DOIprefix\doi{https://doi.org/10.1016/j.epidem.2016.01.002}.
\bibitem[{Wang et~al.(2012)Wang, Wu and Yang}]{Wang2012}
\bibinfo{author}{Wang, X.S.}, \bibinfo{author}{Wu, J.}, \bibinfo{author}{Yang,
  Y.}, \bibinfo{year}{2012}.
\newblock \bibinfo{title}{Richards model revisited: Validation by and
  application to infection dynamics}.
\newblock \bibinfo{journal}{Journal of Theoretical Biology}
  \bibinfo{volume}{313}, \bibinfo{pages}{12 -- 19}.
\newblock \DOIprefix\doi{https://doi.org/10.1016/j.jtbi.2012.07.024}.
\bibitem[{Wu et~al.(2020)Wu, Darcet, Wang and Sornette}]{Wu2020}
\bibinfo{author}{Wu, K.}, \bibinfo{author}{Darcet, D.}, \bibinfo{author}{Wang,
  Q.}, \bibinfo{author}{Sornette, D.}, \bibinfo{year}{2020}.
\newblock \bibinfo{title}{Generalized logistic growth modeling of the covid-19
  outbreak: comparing the dynamics in the 29 provinces in china and in the rest
  of the world}.
\newblock \bibinfo{journal}{Nonlinear Dynamics} \bibinfo{volume}{101},
  \bibinfo{pages}{1561--1581}.
\newblock \DOIprefix\doi{10.1007/s11071-020-05862-6}.
\bibitem[{Xu et~al.(2020)Xu, Yu, Chen and Lu}]{Xu2020}
\bibinfo{author}{Xu, C.}, \bibinfo{author}{Yu, Y.}, \bibinfo{author}{Chen, Y.},
  \bibinfo{author}{Lu, Z.}, \bibinfo{year}{2020}.
\newblock \bibinfo{title}{Forecast analysis of the epidemics trend of covid-19
  in the usa by a generalized fractional-order seir model}.
\newblock \bibinfo{journal}{Nonlinear Dynamics} \bibinfo{volume}{101},
  \bibinfo{pages}{1621--1634}.
\newblock \DOIprefix\doi{10.1007/s11071-020-05946-3}.
\bibitem[{Young and Chen(2021)}]{Young2021}
\bibinfo{author}{Young, P.}, \bibinfo{author}{Chen, F.}, \bibinfo{year}{2021}.
\newblock \bibinfo{title}{Monitoring and forecasting the covid-19 epidemic in
  the uk}.
\newblock \bibinfo{journal}{Annual Reviews in Control}
  \DOIprefix\doi{10.1016/j.arcontrol.2021.01.004}.
\bibitem[{Zhao et~al.(2020a)Zhao, Lin, Ran, Musa, Yang, Wang, Lou, Gao, Yang,
  He and Wang}]{ZhaoShi2020}
\bibinfo{author}{Zhao, S.}, \bibinfo{author}{Lin, Q.}, \bibinfo{author}{Ran,
  J.}, \bibinfo{author}{Musa, S.S.}, \bibinfo{author}{Yang, G.},
  \bibinfo{author}{Wang, W.}, \bibinfo{author}{Lou, Y.}, \bibinfo{author}{Gao,
  D.}, \bibinfo{author}{Yang, L.}, \bibinfo{author}{He, D.},
  \bibinfo{author}{Wang, M.H.}, \bibinfo{year}{2020}a.
\newblock \bibinfo{title}{Preliminary estimation of the basic reproduction
  number of novel coronavirus (2019-ncov) in china, from 2019 to 2020: A
  data-driven analysis in the early phase of the outbreak}.
\newblock \bibinfo{journal}{International Journal of Infectious Diseases}
  \bibinfo{volume}{92}, \bibinfo{pages}{214 -- 217}.
\newblock \DOIprefix\doi{https://doi.org/10.1016/j.ijid.2020.01.050}.
\bibitem[{Zhao et~al.(2020b)Zhao, Shou and Wang}]{Zhao2020}
\bibinfo{author}{Zhao, Y.F.}, \bibinfo{author}{Shou, M.H.},
  \bibinfo{author}{Wang, Z.X.}, \bibinfo{year}{2020}b.
\newblock \bibinfo{title}{Prediction of the number of patients infected with
  covid-19 based on rolling grey verhulst models}.
\newblock \bibinfo{journal}{International journal of environmental research and
  public health} \bibinfo{volume}{17}, \bibinfo{pages}{4582}.
\newblock \DOIprefix\doi{10.3390/ijerph17124582}.
\bibitem[{Zhou et~al.(2020)Zhou, Ma, Hong, Su, Ma, He, Jiang, Liu, Shan, Zhu,
  Zhang and Long}]{Zhou2020}
\bibinfo{author}{Zhou, X.}, \bibinfo{author}{Ma, X.}, \bibinfo{author}{Hong,
  N.}, \bibinfo{author}{Su, L.}, \bibinfo{author}{Ma, Y.}, \bibinfo{author}{He,
  J.}, \bibinfo{author}{Jiang, H.}, \bibinfo{author}{Liu, C.},
  \bibinfo{author}{Shan, G.}, \bibinfo{author}{Zhu, W.},
  \bibinfo{author}{Zhang, S.}, \bibinfo{author}{Long, Y.},
  \bibinfo{year}{2020}.
\newblock \bibinfo{title}{Forecasting the worldwide spread of covid-19 based on
  logistic model and seir model}.
\newblock \bibinfo{journal}{medRxiv}
  \DOIprefix\doi{10.1101/2020.03.26.20044289}.
\bibitem[{Zilany et~al.(2009)Zilany, Bruce, Nelson and Carney}]{Zilany2009}
\bibinfo{author}{Zilany, M.}, \bibinfo{author}{Bruce, I.},
  \bibinfo{author}{Nelson, P.}, \bibinfo{author}{Carney, L.},
  \bibinfo{year}{2009}.
\newblock \bibinfo{title}{A phenomenological model of the synapse between the
  inner hair cell and auditory nerve: long-term adaptation with power-law
  dynamics}.
\newblock \bibinfo{journal}{Journal of the Acoustical Society of America}
  \bibinfo{volume}{126}, \bibinfo{pages}{2390--2412}.
\newblock \DOIprefix\doi{10.1121/1.3238250}.

\end{thebibliography}


%
%
%

\end{document}